\documentclass[12pt,preprint]{aastex}
\begin{document}

\title{The 2011 Eruption of the Recurrent Nova T Pyxidis; the Discovery, the Pre-eruption Rise, the Pre-eruption Orbital Period, and the Reason for the Long Delay}
\author{Bradley E. Schaefer, Arlo U. Landolt\affil{Physics and Astronomy, Louisiana State University, Baton Rouge, LA 70803}}
\author{Michael Linnolt\affil{American Association of Variable Star Observers, 49 Bay State Road, Cambridge MA 02138}}
\author{Rod Stubbings\affil{Tetoora Observatory, Tetoora Road, Victoria, Australia}}
\author{Grzegorz Pojmanski\affil{Warsaw University Observatory, Al. Ujazdowskie 4, 00-478
Warszawa, Poland}}
\author{Alan Plummer\affil{Variable Stars South, Linden Observatory, 105 Glossop Road, Linden, NSW, Australia }}
\author{Stephen Kerr\affil{American Association of Variable Star Observers, Variable Stars South, Astronomical Association of Queensland, 22 Green Avenue, Glenlee, Queensland, Australia}}
\author{Peter Nelson\affil{Ellinbank Observatory, 1105 Hazeldean Road, Ellinbank 3821, Victoria, Australia }}
\author{Rolf Carstens\affil{American Association of Variable Star Observers, Variable Stars South, CBA,
Geyserland Observatory, 120 Homedale Street, Rotorua 3015, New Zealand}}
\author{Margaret Streamer\affil{Lexy's Palace Observatory, 3 Lupin Place, Murrumbateman, NSW, Australia}}
\author{Tom Richards\affil{Variable Stars South, Pretty Hill Observatory, PO Box 323, Kangaroo Ground 3097, Victoria, Australia }}
\author{Gordon Myers\affil{Center for Backyard Astrophysics, Columbia University, 538 W. 120th, New York, NY 10027}}
\author{William G. Dillon\affil{American Association of Variable Star Observers, 4703 Birkenhead Circle, Missouri City, TX 77459}}


\begin{abstract}

We report the discovery by M. Linnolt on JD 2455665.7931 (UT 2011 April 14.29) of the sixth eruption of the recurrent nova T Pyxidis.  This discovery was made just as the initial fast rise was starting, so with fast notification and response by observers worldwide, the entire initial rise was covered (the first for any nova), and with high time resolution in three filters.  The speed of the rise peaked at 9 mag/day, while the light curve is well fit over only the first two days by a model with a uniformly expanding sphere.  We also report the discovery by R. Stubbings of a pre-eruption rise starting 18 days before the eruption, peaking 1.1 mag brighter than its long-time average, and then fading back towards quiescence 4 days before the eruption.  This unique and mysterious behavior is only the fourth known (with V1500 Cyg, V533 Her, and T CrB) anticipatory rise closely spaced before a nova eruption.  We present 19 timings of photometric minima from 1986 to February 2011, where the orbital period is fast increasing with $P/\dot{P}=+313,000$ years.  From 2008-2011, T Pyx had a small change in this rate of increase, so that the orbital period at the time of eruption was 0.07622950$\pm$0.00000008 days.  This strong and steady increase of the orbital period can only come from mass transfer, for which we calculate a rate of $1.7-3.5\times10^{-7}$ M$_{\odot}$ yr$^{-1}$.  We report 6116 magnitudes between 1890 and 2011, for an average $B=15.59\pm0.01$ from 1967-2011, which allows for an eruption in 2011 if the blue flux is nearly proportional to the accretion rate.  The ultraviolet-optical-infrared spectral energy distribution is well fit by a power law with $f_\nu \propto \nu^{1.0}$, although the narrow ultraviolet region has a tilt with a fit of $f_\nu \propto \nu^{1/3}$.  We prove that most of the T Pyx light is {\it not} coming from a disk, or any superposition of blackbodies, but rather is coming from some nonthermal source.  We confirm the extinction measure from {\it IUE} with $E(B-V)=0.25\pm0.02$ mag, although we find problems with all prior distance determinations and are left only with $1000 \lesssim D \lesssim 10000$ pc.

\end{abstract}

\section{Introduction}

T Pyxidis (T Pyx) was the first discovered recurrent nova (RN) with three known eruptions, and this over a dozen years before any other RN had even their second discovered eruption (Schaefer 2010).  T Pyx is one of the quintessential RN, with six eruptions in 1890, 1902, 1920, 1944, 1967, and now 2011.  Over the past few decades, the quiescent magnitude of T Pyx has been near V=15.5, while peaking at V=6.4 (Schaefer 2010).  T Pyx is slow among RNe, with its overall rise taking 40 days and the decline by three magnitudes from the peak ($t_3$) taking 62 days (Schaefer 2010).  T Pyx is one of two RNe with an orbital period (P=0.076 days; Schaefer et al. 1992; Patterson et al. 1998; Uthas et al. 2010) that is inside or below the period gap.  T Pyx is the only RN that has a nova shell, and this shell has a surprisingly slow expansion velocity plus a structure of thousands of discrete knots (Duerbeck \& Seitter 1979; Williams 1982; Shara et al. 1989; 1997).  The homologous expansion of the individual shell fragments proves that they suffer no significant deceleration, and the expansion velocities, viewed over 13 years by {\it Hubble Space Telescope}, prove that the nova shell was ejected in the year 1866$\pm$5 at a velocity of 500-715 km s$^{-1}$ (Schaefer et al. 2010a).  This velocity is greatly too low to be from a RN event, and with the measured mass in the shell ($\sim$10$^{-4.5}$ M$_{\odot}$) being much greater than is possible for a RN event, we know that the 1866 event must have been a normal classical nova eruption with a very long time of prior quiescence (Schaefer et al. 2010a).

A key and unique fact about T Pyx is that its quiescent brightness has been secularly fading from 1890 to 2011 (Schaefer 2005).  Before the 1890 eruption, archival plates show T Pyx to be constant at B=13.8, the next four inter-eruption intervals have B-band magnitudes of 14.38, 14.74, 14.88, and 14.72, while after the 1967 eruption the quiescent B magnitude fades from 15.3 (around 1968) to 15.5 and fainter (from the 1980's until 2011).  This secular decline is fundamental for understanding the evolution and fate of T Pyx, as well as the timing of the eruptions.

T Pyx has long been a perplexing system, because there was no known way to combine a short orbital period and a high accretion rate into one system.  (The accretion rate must be high, $\sim$10$^{-7}$ M$_{\odot}$ yr$^{-1}$, so as to provide enough mass to trigger eruptions with a typical recurrence time interval of two decades.)  That is, a binary with a period below the gap should have angular momentum loss only from gravitational radiation, and this will lead to an accretion rate many orders of magnitude smaller than required (Patterson 1984).  This dilemma was solved by Knigge et al. (2000) who realized that the high accretion rate could be powered by a hot luminous source on the white dwarf that is irradiating the companion star.  For T Pyx, this idea was extended by Schaefer et al. (2010a), where the classic nova eruption of 1866 started a high accretion rate which drives a largely self-sustained supersoft source (with nuclear burning on the surface of the white dwarf).  This mechanism can only work for binaries with short orbital periods and where a substantial magnetic field channels the material into a small area.  The secular fading demonstrates that the supersoft source is not fully self-sustaining.  Schaefer \& Collazzi (2010) generalizes this situation to a whole class of stars, called V1500 Cyg stars, which suffer this same secular fading after their eruptions, and which also have short orbital periods and highly magnetized white dwarf stars.  So the picture we have for T Pyx is that it started out as an ordinary cataclysmic variable with a long duration in quiescence, then an ordinary nova eruption in 1866, which ignited a supersoft X-ray source whose irradiation drives a high accretion rate and hence recurrent nova eruptions.  Over time, the accretion rate declines and the quiescent brightness fades.

By far the most important question relating to RNe is whether they are the progenitors of Type Ia supernovae.  RNe are one of the best progenitor candidate classes, because they certainly have white dwarfs near the Chandrasekhar mass being fed with a very high accretion rate.  Nevertheless, T Pyx (and any short-period RNe) cannot be a Type Ia progenitor (Schaefer et al. 2010a; Schaefer 2010).  The reason is that the ordinary nova eruption in 1866 easily dominates the dynamics of the system, and ordinary nova eruptions eject much more mass than has been accreted by the white dwarf, so the white dwarf is actually loosing mass and will not become a supernova.  Another way of saying this is that the secular fading shows that the interval during which T Pyx is a RN is short, perhaps under two centuries, and thus will have little effect even if the RN event causes a net gain of mass on the white dwarf.  The long-period RNe remain as strong progenitor candidates (due to their high accretion rate being driven by the evolution of the donor star so as to be sustained and long in duration), but the short-period RNe cannot sustain the high accretion and so are not supernova progenitors.

The secular fading is also critical for understanding the timing and triggering of the RN events.  Schaefer (2005) demonstrated that the inter-eruption intervals during which the quiescent magnitude was faint on average lead to a long time between eruptions, while a bright interval lead to a short time between eruptions. The physics is simple, with the amount of accreted mass required to trigger an eruption being constant for any one RN, with the accretion rate being given by the blue flux dominated by the accretion disk, so that a high quiescent brightness implies a high accretion rate and a short interval needed to accrete the trigger mass.  With calibration from prior inter-eruption intervals, it becomes possible to predict the approximate date of upcoming eruptions.  As such, Schaefer (2005) predicted that U Sco would erupt in 2009.3$\pm$1.0, and this advance notice allowed for an incredible coverage of the eruption in 2010.1, so that this now is the all-time best observed nova event (Schaefer et al. 2010b).    Schaefer (2005) also predicted that the next eruption of T Pyx would be in 2052$\pm$3, while Schaefer et al. (2010a) extended this to predict that the next eruption will be after the year 2225.   This prediction was shown to be horribly wrong by the T Pyx eruption in 2011, so the issue now is to understand where the prediction went wrong.

\section{Discovery of the 2011 Eruption}

Starting with the 1890 eruption, the inter-eruption intervals were 11.9, 17.9, 24.6, and 22.1 years (Schaefer 2010).  In the middle 1980's many people around the world (including many of the authors of this paper) realized that T Pyx `should' go off soon.  The simple logic was that the prior inter-eruption interval was 22 years, and 22 years after the 1967 eruption placed the expected eruption around 1989.  Due to this, many of the world's best observers kept T Pyx under nightly watch.

As the 1990's stretched on, the expected eruption did not occur.  Vigilance was maintained and occasional public exhortations were published in many journals.  Any eruption after 1967 would certainly have been discovered because the T Pyx eruption duration is two-thirds of a year and so no eruption would have been missed due to observational gaps.  By the year 2005, the interval since the prior eruption extended to 38 years, with no explanation for this long delay.

Schaefer (2005) provided the explanation.  The key point was that T Pyx has been declining overall in brightness ever since before the 1890 eruption, and this means that the accretion rate in the system was declining, so that it would take longer to accumulate the trigger mass on the surface of the white dwarf.  In particular, from the 1944 to the 1967 eruption the average B-band magnitude was 14.72$\pm$0.03, while from 1967 to 2005 the average B-band magnitude was 15.49$\pm$0.01 mag.  The blue flux from T Pyx is entirely dominated by the accretion light, so the B-band flux is determined by the accretion rate.  The dimming by 0.77 mag (a factor of two drop in accretion luminosity) requires a fall in the accretion rate by at least a factor of two, and this forces an increase in the inter-eruption time interval.  The long delay since 1967 was simply because the rate of accumulation on the white dwarf surface had at least halved.

With the prediction from Schaefer (2005) that the next eruption was a long way in the future, professional preparations for the next eruption (like the long-running Target of Opportunity program with the {\it Hubble Space Telescope}) were stopped.  But this did not deter many amateur observers (who were generally fully aware of the prediction) from continuing to regularly check for eruptions, report visual magnitudes, and make long photometric time series with CCDs.  For example, Stubbings started in early 1998 in making 2004 visual estimates of the brightness of T Pyx with a 16-inch reflecting telescope in Victoria Australia, while Richards made CCD time series observations on five nights in early 2008 with 661 brightness measures.

From 2008 until 31 March 2011 (JD 2455652), T Pyx held at a steady magnitude of V=15.5, with the usual periodic variations of $\lesssim$0.2 mag.  For the 610 observations in the last 4000 days before March 2011, Stubbings never saw T Pyx brighter than V=15.0.  Thus, it was a distinct surprise when Stubbings saw T Pyx at V=14.5 on JD2455657.0104 (UT 2011 April 5.51), with the realization that this could be an initial rise of the next T Pyx eruption.  With this exciting possibility, Stubbings checked T Pyx five more times before dawn, but saw only variations from 14.4 to 14.7.  On the next night T Pyx was slightly fainter, and it continued a slow fade until JD2455662.0417 (UT 2011 April 10.54).  The light curve just before the sharp initial rise to the eruption shows a significant short-duration pre-eruption rise, which started roughly 18 days before the eruption, peaked at around V=14.4 mag (close to 1.1 mag above the prior quiescent level), and declined towards quiescence before the eruption.

The existence of this pre-eruption rise has been confirmed with the archival data from the {\it All Sky Automated Survey} (ASAS, Pojmanski 2002).   For the year prior to JD 2455654.5972, T Pyx was seen to vary about its average with an RMS scatter of 0.14 mag.  On JD 2455656.62097 (0.4 days before Stubbings discovery of the rise), T Pyx was 0.43 mag brighter than the average, with this being the brightest magnitude by far in the prior year.  This is the confirmation of the pre-eruption rise.  The apparent duration of the pre-eruption rise, as seen with the ASAS data alone is roughly one week, or perhaps a bit longer.

This pre-eruption event is one of an unusual phenomena that is still completely mysterious.  Collazzi et al. (2009) report an extensive survey of archival data for 22 novae with well-sampled pre-eruption light curves, which showed that three of these (V1500 Cyg, T CrB, and V533 Her) have pre-eruption rises with greatly different time scales and morphologies.  Table 1 summarizes the properties of these events, and both the rises and the nova systems are a rather wide range of properties.  In all cases, the pre-eruption rise cannot be some early part of the usual thermonuclear runaway because the anticipation times and the durations (weeks and years) are greatly longer than both  the theoretical runaway timescale (hours) and the observed rise  times (hours).  In all cases, the pre-eruption event is highly significant and unprecedented for many decade intervals before the eruptions and after the eruptions.  For the example of V1500 Cyg, the fast apparently-exponential rise was detected at high significance on five plates, getting as bright as B=13.5 mag (i.e., 7 mag brighter than quiescence), with no other such brightenings from 1890 to 2013.  The uniqueness and the time coincidence immediately before the eruption provide proof of a causal connection.  It is difficult to understand this connection, because the pre-eruption brightness level is dominated by the accretion which is governed by the mass falling off the donor star, while the timing of the eruption is governed by conditions at the bottom of the accretion layer deep under the surface of the white dwarf.  How could the donor star near the L1 point possibly {\it anticipate} that conditions deep in the white dwarf will soon heat to the trigger threshold?  We regard the explanation of these pre-eruption phenomena as a premier challenge for nova theory.  Presumably, the durations, start times, and luminosities will provide sensitive indicators of conditions at the time of the nova trigger.

Linnolt had also been monitoring T Pyx since March 2009, waiting for an eruption, as part of a larger program of monitoring recurrent novae for eruptions (e.g., Schaefer et al. 2010b).  On JD2455664.8590 (UT 2011 April 13.36), Linnolt observed T Pyx at V=14.5 and realized that this was unusually bright, and hence a possible indication of an eruption.  With anticipation, Linnolt checked T Pyx early the next night (JD2455665.7931, UT 2011 April 14.29), saw that the star was bright (V=13.0), and immediately realized that an eruption had started.  Within 15 minutes, the discovery was announced on the AAVSO discussion group which is closely monitored by interested amateurs and professionals worldwide.  The discovery was quickly confirmed by Australian visual observers Plummer (JD2455665.8847 V=12.2) and Kerr (JD2455665.941 V=11.3).  CCD observers Nelson, Carstens, and Streamer started taking time series photometry in multiple bands by JD2455665.9973.  

Serendipitously, Schaefer actually took the first images of T Pyx certainly in eruption, 7.6 hours before Linnolt's discovery.  Fortuitously, on that night, Schaefer had half a night of service observing on the SMARTS 0.9-m telescope on Cerro Tololo in Chile, with the bulk of the time being used to get a long time series covering an eclipse of the recurrent nova U Sco (see Schaefer 2011).  This was scheduled many months in advance, and the night was filled out with images of other RNe, including two B-band images of T Pyx.  In the excitement and activity of the T Pyx eruption, these two images were overlooked for several months, until the U Sco data were analyzed in due course.  The two 100-second exposures (JD 2455665.4752 and 2455665.4986) show T Pyx at B=12.51$\pm$0.01 and B=12.50$\pm$0.01 respectively, which is over three magnitudes brighter than in quiescence, so the fast rise was certainly well underway.  Intriguingly, this magnitude (B=12.50) is brighter than the discovery magnitude of Linnolt (V=13.0) made 7.6 hours later.  One possibility is that T Pyx had a stutter in its rise, with some local maximum in its light curve hours before discovery, so that theorists would have to explain why there should be a significant {\it drop} in brightness near the start of the fast initial rise.  Another possibility is that the initial light from the rise was extremely blue so that the V-band initial rise could have been monotonic and steady.  Interpolation between Linnolt's pre-discovery and discovery magnitudes to the time of the SMARTS image pair gives V=13.51 mag, and hence a required $B-V= -1.0$ mag for the extreme color hypothesis.  The two possibilities (a non-monotonic rise or $B-V= -1.0$) can be used to test theoretical model.

To the best of our knowledge (e.g., Payne-Gaposchkin 1964; Robinson 1975; Collazzi et al. 2009; Strope, Schaefer, \& Henden 2010; Schaefer 2010), the 2011 eruption of T Pyx has by far the best observed initial rise from quiescence.  Only a handful of old novae have had just a few points taken during the later half of their initial fast rise, whereas T Pyx has high time resolution time series with near full-time coverage and in multiple bands.  Indeed, T Pyx is the only nova with any observations in the middle or the first part of the initial fast rise.  The very fast alert to the world and the very fast reaction by observers worldwide made this possible.

\section{Our New Observations}

We have collected a wide variety of new observations presented here for the first time:

\subsection{Pre-eruption Photometry of Quiescence}

Schaefer has returned to the Harvard College Observatory and remeasured the many archival photographic plates, with this providing the photometric history of T Pyx from 1890 to 1953 and the basis for knowing that T Pyx has suffered a severe decline in quiescent brightness from 1890 to present.  The brightness of T Pyx was made by direct comparison with nearby comparison stars with magnitudes in the B magnitude system (Schaefer 2010).  The spectral sensitivity of the Harvard plates is identical to that of the B magnitude system (indeed, these plates were involved in the original definition of the system), and explicit measures of the color terms (corrections proportional to $B-V$) always show that these are closely equal to zero.  With this, the returned magnitudes are exactly in the B-magnitude system.  Repeated experiments with the Harvard plates show that an average plate (with a good nearby comparison sequence) will produce a magnitude with a one-sigma uncertainty of 0.15 mag.  Of particular interest are the first two magnitudes from Harvard, which are {\it before} the 1890 eruption.  The 1890 eruption was recorded on a Harvard plate at B=8.05 at a time 25 days after the two earliest plates, so we have to wonder whether we have a record of some pre-eruption event or the initial rise (rather than a record of T Pyx in quiescence).   We have two plates separated by 2 days both showing the same magnitude (B=13.80), so we know that these plates could not show the initial rise.  The 2011 eruption light curve has B=8.05 around 1 day and 70 days after the start of the initial fast rise, so the two plates at B=13.80 taken 25 days earlier cannot be part of either the initial fast rise or a pre-eruption bump.  Therefore, these two plates show the pre-eruption normal quiescent phase of  T Pyx.

Starting in 1996, Stubbings began a monitoring program on T Pyx, accumulating 2004 visual estimates of T Pyx.  In his 15 years of seeking the next eruption, T Pyx was checked more frequently than once every two nights, all visually with the AAVSO sequence and a 16-inch Newtonian reflector telescope from Tetoora Road Observatory.  

Starting in early 2003, Dillon began a long-term monitoring campaign on T Pyx, accumulating 125 magnitude estimates before the start of the eruption.   These magnitudes were taken with a CCD camera through a V filter with the AAVSO sequence of comparison stars.

Starting in March 2009, Linnolt included T Pyx in his program of regularly checking for eruptions of recurrent novae.  His observations were from several sites on different islands in the Hawaiian Islands.  He used a 20-inch aperture Newtonian reflector telescope operating visually at 500$\times$ magnification, so that observations down to V=17 were possible.  Before he obtained his 20-inch telescope, his use of a smaller telescope allowed him to see T Pyx crowded by the nearby star, so he reported only an upper limit because he could not confidently resolve T Pyx as a separate star.

From 2005-20011, Schaefer has used the SMARTS telescopes on Cerro Tololo to monitor the slowly fading quiescent magnitude of T Pyx.  From 2005-2009, BVRI magnitudes were measured with CCD images taken with the SMARTS 1.3-m telescope.  There was a hiatus in 2010 due to all the observational resources being focused on the eruption of U Sco.  The recommencement of monitoring in 2011, made with the SMARTS 0.9-m telescope, fortuitously caught the first rise of the T Pyx eruption. 

From July 2010 until the eruption, ASAS covered T Pyx frequently as a regular part of its sky survey.  ASAS uses a  telescope with a 200-mm diameter f/2.8 lens located at the Las Campanas Observatory in northern Chile (Pojmanski 2002).   The images were taken through a Bessel V filter and magnitudes are tied to a large number of comparison stars in the field with magnitudes in the Tycho catalog. This procedure can lead to moderate offsets for an object with unusual color (like T Pyx).  Nevertheless, the differential photometry is excellent to within the statistical uncertainties (typically 0.09 mag).  From this, we have 40 V-band magnitudes, critically including a confirmation of the pre-eruption rise.

All our pre-eruption photometry are presented in Table 2.  The first column gives the Julian Date of the observation (or the middle of the CCD exposure), the second column gives the band, the third column gives the magnitude and its one-sigma error bar, while the last column gives the source of the magnitude.  We have also included some scattered observations from the AAVSO data base, with the observer being indicated by their three-letter observer code in parentheses.  We have also included a variety of published T Pyx magnitudes in quiescence, primarily for the convenience of readers so that they do not have to scramble around the literature.  The result is that Table 2 contains 3850 magnitudes for T Pyx in quiescence, and this represents essentially all the useable quiescent light curve from 1890 to 2011.  The total number of magnitudes in Table 2 (including quiescence, the pre-eruption rise, and the initial rise until JD 2455668.0) is 6116.  The entire table is available only in the on-line electronic version, while the printed paper has only the first ten lines to show format, plus the middle lines that cover the critical pre-eruption rise and the initial fast rise.  Table 5 of Schaefer (2010) has 603 magnitudes of T Pyx in eruption from 1967 and earlier.

The entire B-band quiescent light curve from 1890 to 2011 is plotted in Figure 1, with the vertical lines indicating the times of eruptions.  Importantly, we see a steady decline in quiescent brightness from before the 1890 eruption (with B=13.8) up until the 2011 eruption (with B=15.7).  With the B-band brightness depending entirely on the accretion light in the system, we see that the accretion rate has been falling off dramatically over the 122 year interval.  This falloff is critical for understanding the changing time between eruptions (Schaefer 2005; Section 5.4), deriving the evolutionary status of T Pyx (Schaefer et al. 2010a), and for knowing that T Pyx will not become a Type Ia supernova (Schaefer et al. 2010a).

The B-band light curve of T Pyx in quiescence from 1967 to 2011 is plotted in Figure 2.  The earliest points show T Pyx somewhat above quiescence in the tail of the 1967 eruption.  When photometry picks up in 1976, we see a flat light curve, with the usual flickering superposed.  From 1976 to 2004.5, the average B magnitude is 15.50, with a formal one-sigma error bar (the observed RMS scatter divided by the square root of the number of observations) of 0.01 mag.  After 2004.5, the quiescent brightness appears to have taken a step fainter, with this same effect visible in the V-band light curve.  From the middle of 2004 until early 2011, the average is $B=15.68\pm0.01$.  We know of no special event in 2004, but this appears to be just a normal part of the star fading since 1890.

We have used our full table of quiescent magnitudes to determine the average B-band magnitude between eruptions for 6 inter-eruption intervals from before the 1890 eruption up to the 1967-2011 interval.  The V magnitudes are converted to B magnitudes by using the long term average of $B-V=+0.12$ (Schaefer 2010).  (The R and I magnitudes are few in number and usually nearly simultaneous with a B and V magnitude, so the R and I magnitudes were not used.)  Nightly averages are formed, so that a few nights with many magnitudes as part of a time series (which might or might not be representative of the whole interval) do not dominate the average.  Later in this paper, we will be needing an average of the B-band flux (not magnitude), so we have defined a unit of flux equal to the flux when T Pyx is at B=16.00 mag.  That is, $F_B=10^{-0.4(B-16)}$.  These fluxes are then computed for each nightly average, and then combined as an average over each inter-eruption interval to give $\langle F_B \rangle$.  The uncertainty for each average is taken to be the RMS scatter divided by the square root of the number of nights.  (If the nightly averages are first formed into yearly averages and these are averaged, then we get fluxes which are essentially identical.)  Table 3 gives the pre-eruption intervals, their durations ($\Delta T$ from Schaefer 2010), the average B-band magnitudes over each interval ($\langle B \rangle$), the average $F_B$ for those intervals, the values of $\langle F_B \rangle\Delta T$ (which should be nearly constant), and the values of $\langle F_B^{1.13} \rangle\Delta T$.  This last column will be discussed in Section 5.4.

\subsection{Pre-eruption Minima Times}

On five nights in January 2008, Richards made long time series photometry on T Pyx, all with 100-second integrations on a 0.41-m Ritchey-Chretein reflecting telescope at Pretty Hill Observatory near Melbourne Australia.  The images were all taken with an unfiltered CCD, with the T Pyx magnitude being taken by differential photometry with respect to the V-band magnitudes of the comparison stars (designated as `CV' band), so the effective magnitudes are similar to the V-band but nevertheless different.  This non-standard band is perfect for getting the shape of the light curve, as needed to time the minima.

On three nights in February 2011, Myers made long time series photometry of T Pyx.  The reason for this photometry, and for the nearly contemporaneous time series of Nelson, was as a support for observations of T Pyx with the {\it Chandra} X-ray satellite (Balman 2012).  The CCD images were made without a filter (in the `CV' band) with 120-second exposures, with readout times of 64 seconds, with a 0.25-m telescope.  The observing site is in Mayhill New Mexico.

On five nights in February 2011, Nelson made long time series photometry on T Pyx using a 0.32-m corrected Dall-Kirkham telescope at Ellinbank Observatory, in Victoria Australia.  A total of 192 magnitudes were made with an unfiltered CCD and V magnitudes for the comparison stars (hence, the `CV' band).  The integration times were always 120 seconds, resulting in an average time between exposures of 170 seconds.  

The magnitudes from these pre-eruption time series are included in Table 2.  The magnitudes were grouped together into five intervals, each from a single observer, and these groups were then used to measure the times of minimum in T Pyx's light curve.  We have performed chi-square fits of the folded light curve to a template.  The folded light curve is folded around the period from the parabolic ephemeris of Equation 1 (see below).  (We are dealing with time series of such short length that any approximate period would have been adequate.)  The template we used was a flat light curve from phases 0.25-0.75 with a linear and symmetric decline to a minimum at zero phase of 0.15 mag fainter (see Figure 10 of Patterson et al. 1998).  The derived times of minima were then corrected to heliocentric Julian Dates (HJD).  This procedure yields a formal uncertainty on the time of minimum, but this only accounts for the measurement errors.  The uncertainty is actually dominated by the orbit to orbit variations in the light curve (see Figure 54 of Schaefer 2010 and Figure 8 of Patterson et al. 1998) caused by the usual flickering plus other unstable periodicities.  When one orbit is considered, the light curve occasionally does not have an obvious minimum.  However, when many orbits are considered together, the combined light curve always looks similar to the template.  So the accuracy of the times of minima should scale as the inverse square root of the total duration of data in the time series in terms of the orbital period.  For a group of similar observations over a relatively small time range, the scale of this uncertainty can be set by looking at the scatter in the O-C curve about some smooth best fit line.  So for example, the uncertainty in the minimum times reported by Patterson et al. (1998) can be calculated as $0.0033 (\Delta/0.076)^{-0.5}$ days, with $\Delta$ being the total duration of the observation (not necessarily contiguous) in days.  With this, Table 4 presents the available times of minimum (in heliocentric Julian days) with their uncertainties.  Included are six times of minimum from Patterson et al. (1998).

\subsection{Photometry of the Initial Rise}

The pre-eruption rise was covered by the visual observations of Stubbings and the CCD observations of ASAS, as described in Section 3.1.  The initial fast rise was recorded by the SMARTS CCD images of Schaefer and the visual discovery observations of Linnolt, as described in Section 3.1.  A blow up of the light curve around the time of the pre-eruption rise and the initial fast rise is displayed in Figure 3.  

Immediately after the announcement of the discovery by Linnolt, we began an intensive set of time series images in the B, V, R, and I bands.  This intensive set of observations has continued all the way through the peak, and indeed continues with daily long time series until the time of this writing (February 2013).  Plummer, Kerr, and Stubbings provided immediate visual confirmation of Linnolt's discovery.  Nelson, Carstens, and Streamer immediately began all-night fast time series photometry in the V-band.  These time series provide a unique record of the fast initial rise from the earliest time, for direct comparison with theory.  Streamer also took parallel time series in B, V, and I, so we have a totally unique complete record of the color development of the fast initial rise from the earliest times.  Further observations have been taken from the archives of the AAVSO.  In this paper, we will only cover the the pre-eruption quiescence, the pre-eruption rise and the first 2.2 days of the eruption.  All these magnitudes are recorded in Table 2.  Figure 4 shows the light curve of the initial fast rise.

We have large numbers of observations of time series in multiple colors, where pairs of images through different filters were taken within seconds of each other, and so we can easily calculate the T Pyx colors at many times throughout the fast initial rise.  The first measures of the color (when T Pyx was around eleventh magnitude) have median $B-V= -0.02$, but this slowly changes to $B-V=+0.23$ by two days after the start.  At this time, just as the light curve is nearly flat, the $B-V$ color curve also goes nearly flat.  Just after discovery (around JD2455666.07), the median colors are $B-V=-0.02$ and $V-I=+0.39$.  These colors are inconsistent with a blackbody, as for example an A0 star with a temperature near 10,000K ($B-V=0.00$) has the V-I color equal to 0.00, which is substantially different than that seen in T Pyx.  Half a day after discovery (JD2455666.65), the median colors are $B-V=+0.06$, $V-R=+0.27$, and $V-I=+0.37$.  At the time when the fast initial rise breaks (JD2455668.0), the median colors are $B-V=+0.23$ and $V-I=+0.58$.  Figure 5 shows our derived B-V color curve through the initial fast rise.

\subsection{{\it GALEX} Spectrum and the Spectral Energy Distribution}

We have obtained an ultraviolet spectrum of T Pyx in quiescence with the {\it GALEX} satellite.  This is from {\it GALEX} Cycle 2, with Schaefer as Principal Investigator.  The observation time was 2005 December 20 starting at UT 03:30:05 (JD 2453724.646), with an effective total exposure of 2503 seconds.  {\it GALEX} gives a grism spectrum with a FUV band covering 1344-1786 \AA,  and a NUV band covering 1771-2831 \AA.  We have used the standard data analysis pipeline, which returns a fully calibrated combined spectrum with the spectral flux, $f_{\lambda}$, in units of erg s$^{-1}$ cm$^{-2}$ \AA$^{-1}$.  The FUV magnitude is 15.81$\pm$0.01 and the NUV magnitude is 16.00$\pm$0.01.  The resultant spectrum is of good quality, although the noise at the edges of the bands gets fairly large.  

The {\it GALEX} spectrum a very blue continuum with no blatant emission features, plus a prominent broad dip centered on 2175\AA.  We have dereddened the spectrum by fitting for the extinction that best removes the 2175\AA bump, for which we find $E(B-V) = 0.239\pm0.004$ mag.  This spectrum, with moderate binning plus clipping of several end points with very large error bars, is displayed in Figure 6.  We see a smooth continuum, plus only one significant spectral line (the C IV line at 1550\AA).  This spectrum is well fit by a power law, for which we find an index of -2.25$\pm$0.03, while the normalization has a formal error of 1.1\%.  Our best fit is  $f_{\lambda}=2.78\times 10^{-6} \lambda ^{-2.25}$ erg s$^{-1}$ cm$^{-2}$ \AA$^{-1}$, for wavelength $\lambda$ measured in Angstroms.  For a broad spectral energy distribution (SED), it is usual to represent the spectral flux as $f_{\nu}$ (in units of Jansky, 1 Jy = $10^{-23}$ erg s$^{-1}$ cm$^{-2}$ Hz$^{-1}$), where we have the relation $\nu f_{\nu}=\lambda f_{\lambda}$.  In a plot of $f_{\nu}$ versus frequency ${\nu}$, our best fit spectral slope is $f_{\nu} \propto \nu ^ {0.25}$  This is essentially identical to the best-fit power law from the average {\it IUE} data (Gilmozzi \& Selvelli 2007). 

We have also constructed a SED for T Pyx from 1456\AA~to 22,200\AA~by using the {\it GALEX} spectrum along with optical and infrared photometry.  Schaefer (2010) measured the UBVRIJHK colors of T Pyx on several occasions contemporaneous with the {\it GALEX} observations and has converted these magnitudes into $f_{\nu}$ with extinction corrections.  (The original paper had already pointed out the U-band $f_{\nu}$ as being a likely error, and we have now recognized that the U-V color for T Pyx had been incorrectly transferred from Table 25 to Table 30 of Schaefer 2010 without the negative sign, and so the SED in Figures 70 and 71 show a false turn down in the ultraviolet. This error is now corrected in this paper.)  To display the SED, we plot logarithmic scales with frequency, $\nu$, on the horizontal axis, and $\nu f_{\nu}$ on the vertical axis, in part because this presentation shows the peaks where the most energy is coming out.  For T Pyx, our SED is plotted in Figure 7.  We see a rising spectrum towards the ultraviolet, so most of the T Pyx accretion energy is coming out for wavelengths shorter than 1456\AA.  This will not surprise anyone, but it does illustrate that we are missing most of the accretion light.  The overall SED rises remarkably close to a simple power law, even though we can see deviations, for example where the slope is different in the ultraviolet.  Our best fit power law for $f_{\nu}$ has an index of essentially unity, so we have $f_{\nu} = 0.009$Jy$(\nu/10^{15}$Hz).  Gilmozzi \& Selvelli had a similar result with $f_{\lambda} \propto \lambda^{-2.9}$ or  $f_{\nu} \propto \nu^{0.9}$.  This is completely different from a hot blackbody spectrum, where the Rayleigh-Jeans region has $f_{\nu} \propto \nu^2$.

We can also place an X-ray point onto the SED.  Balman (2012) reports that the central source has a 0.2-9.0 keV flux of $1.0-5.8 \times 10^{-14}$ erg s$^{-1}$ cm$^{-2}$.  This translates into something like $10^{-9}$ Jy for a frequency around $2\times10^{17}$ Hz, so $\nu f_{\nu}$ is around $2\times10^{8}$ Hz Jy.  This point is half-again off the right side of the box in Figure 7, and far below the bottom.  So we know that the power law rise so prominent in Figure 7 must break sharply downwards somewhere between $2\times10^{15}$ Hz and $2\times10^{17}$ Hz.

This SED will give the accretion luminosity of T Pyx as $L=4\pi D^2 \int f_{\nu} d\nu$.  For a fiducial distance of 3000 pc, the spectral range from $1-20 \times 10^{14}$ Hz gives a luminosity of $2 \times 10^{35}$ erg s$^{-1}$.  If the observed power law rise to high photon energies continues until a sharp break (as constrained by the X-ray observation), for a total range in frequency of $1-2000 \times 10^{14}$ Hz, we get a maximal accretion luminosity of $2 \times 10^{39}$ erg s$^{-1}$.  Even for a set fiducial distance, this allows a rather wide range of possible accretion luminosities.

\section{Measuring Properties of the T Pyx System}

We have used our observations, plus results from the literature, to derive the properties of the T Pyx system:

\subsection{The Steady Period Change $\dot{P}$}

In the classic review of RNe, Webbink et al. (1987) concluded that T Pyx must have a relatively short orbital period (like those of the typical classical novae) due to the lack of any red excess.  Early photometric periods suggested were 0.069 days (Szkody \& Feinswog 1988) and 0.1433 days (Barrera \& Vogt 1989).  With a much larger set of time series photometry, Schaefer (1990) found a stable and highly significant modulation but had severe daily alias problems, for which the highest peak in the Fourier transform was at 0.099 days, while another lower peak was at 0.076 days.  Schaefer et al. (1992) collected an even larger set of time series from 1966 to 1990 and determined the period to be 0.07616$\pm$0.00017 days.  Patterson et al. (1998) made long runs from 1996 to 1997, confirmed this orbital period, discovered a steady period change, and discovered a complicated structure of weaker periodicities that vary in amplitude.  Uthas et al. (2010) demonstrated that the stable 0.076 day photometric period equals the orbital period as measured with a radial velocity curve.  They also used many minimum timings from 1996-2009 (from the records of the {\it Center for Backyard Astronomy}) to give a high precision ephemeris that includes a parabolic term.  This gives the minima times (expressed in Julian Dates) as 
\begin{equation}
T_{Uthas} = 2451651.65255 + 0.076227249 N + 2.546 \times 10^{-11} N^2,
\end{equation}  
where N is the number of orbital cycles from the epoch.  The quadratic coefficient has an uncertainty of 2.1\%.

The observed minima times ($T_{min}$) in Table 4 (see Section 3.2) can be plotted by means of an O-C diagram.  For this plot, we compare the observed minus the predicted minimum times (O-C, measured in days), where the predicted times are taken from some linear ephemeris.  We have chosen a linear ephemeris of 
\begin{equation}
T_{linear} = 2451651.65255 + 0.076227249 N ,
\end{equation}  
which is just the Uthas ephemeris without the quadratic term.  The observed-minus-calculated difference is 
\begin{equation}
O-C = T_{min} - T_{linear}.
\end{equation} 
The O-C curve is a plot of $O-C$ versus time $T$.  If the star obeys the linear ephemeris, then the O-C curve will show all the minimum times to have the usual scatter along the horizontal axis.  If the period is different then given in the linear ephemeris, then the O-C curve will have the points follow along a straight line with some non-zero slope.  If the period changes linearly with time, then the slope of the plot will change linearly with time, and this will appear as a parabola in the O-C diagram.  The $N$ and $O-C$ for the linear ephemeris are tabulated for each of our $T_{min}$ in Table 4.  The O-C plot for our 19 $T_{min}$ measures and the linear ephemeris from Equation 2 is given in Figure 8a.  We see a good parabola, which points to a steady period change in the T Pyx orbital period.  This is not surprising, because Uthas et al. used our data for all the pre-1992 times.  Our best fit parabola to Figure 8a returns an ephemeris very close to that of Uthas et al., so we will simply adopt her ephemeris as the best measure of the steady period change in T Pyx.

For later use in equations involving mass transfer, we need the steady period change, $\dot{P}$, cast into a dimensionless form.  This period derivative can be calculated from Equation 1, with the period over each orbit equaling the time difference between successive minima, so $\dot{P}$ equals twice the quadratic coefficient in Equation 1 divided by the orbital period in days.  Thus, $\dot{P}=6.68 \times 10^{-10}$, with an uncertainty of 2.1\%.  A useful way to express this is $P/\dot{P}=313,000$ years, about a third of a million years.

\subsection{Orbital Period Just Before the Eruption $P_{pre}$}

A highly accurate pre-eruption orbital period ($P_{pre}$), together with a post-eruption period, will give the change in orbital period across the 2011 eruption.  This period change will then give the ejected mass with high accuracy and no need for uncertain quantities (like the distance) or dubious assumptions (like the shell filling factors).  For the case of T Pyx (but not for the other RNe), the dynamical friction of the companion star within the nova envelope will also cause a small period change (Livio 1991; Schaefer et al. 2010a) and this effect must also be taken into account.  The ejected mass is critical for knowing whether the RN white dwarf is gaining or losing mass over the eruption cycle, and hence whether the RN will soon become a Type Ia supernova.  Previously, only two other RNe (U Sco and CI Aql) had a measure of the period change across the eruption (Schaefer 2011).  The T Pyx eruption of 2011 provides an opportunity to measure the period change for a third RN.

As the period of T Pyx changes, it is vital to determine the pre-eruption orbital period for the time immediately preceding the eruption.  To good accuracy, Equation 1 should work well to give the pre-eruption orbital period.  With this, for an epoch of N=52662, the pre-eruption orbital period is 0.07622993$\pm$0.00000017 days.  However, this period is based on minima times only until 2009.  We now have the time series in February 2011, up to 48 days before the eruption.  The possible existence of small period changes just before the eruption is critical where we need part-per-million accuracy.

To search for deviations from the parabolic ephemeris, we constructed another O-C curve where the ephemeris was Equation 1.  That is, $O-C=T_{min}-T_{Uthas}$.  The resultant quadratic O-C values are presented in the last column of Table 4, and the O-C diagram is shown in the lower panel of Figure 8.  If the Uthas et al. (2010) model is perfect, then all the observed minimum times should lie along the horizontal axis with $O-C=0$ to within the plotted error bars.  Indeed, we see that the early time series from Schaefer et al. (1992), Patterson et al. (1998), and from early 2008 by Richards all fall close to the axis.  This is to say that the parabolic ephemeris is a good match to the observed times from 1988-2008.  The two times in 1986 and 1987 are roughly 2-sigma and 1-sigma above the line, with a weak suggestion of deviation from Equation 1 before 1988, although we judge this suggestion to be too weak to accept.  However, the two times from February 2011 are both closely consistent and with 3.7-sigma and 5.1-sigma  deviations from the parabolic ephemeris.  These two measures indicate that T Pyx underwent a very small but significant change in period from 2008 to 2011.  

If the O-C curve is linear from 2008 to 2011, then the orbital period (gotten from just our five minimum times in 2008 and 2011) is $P_{pre}$=0.07622916$\pm$0.00000008 days.  But the steady period change is caused by the mass transfer in the system, and we know that this has not stopped (T Pyx was still flickering in early 2011), so the shape of the O-C curve must still be a parabola from 2008-2011.  In all scenarios, $\dot{P}$ is proportional to $\dot{M}$, and from Section 5.3 we have that $\langle F_B \rangle$ is proportional to $\dot{M}$.  For a magnitude difference of 0.18 mag (see Section 3.1 and Figure 2), the blue-band flux ratio is 0.85, so both $\dot{M}$ and $\dot{P}$ should also have that same ratio for the two time intervals.  With this, for the last few years before the 2011 eruption, $\dot{P}=5.66 \times 10^{-10}$, and $P/\dot{P}=366,000$ years.  We only have minima times in 2008 and 2011, so with the derived $\dot{P}$, we have enough to state a quadratic ephemeris for T Pyx in the years just before eruption.  We then fit the times to get 
\begin{equation}
T_{pre-eruption} = 2455665.9962 + 0.07622950 N_{2011} + 2.16 \times 10^{-11} N_{2011}^2,
\end{equation}  
Again, the coefficient of the quadratic term is just $P\dot{P}/2$.  Here, the epoch was chosen to be the time of minimum closest to the start of the eruption.  That is, the start of the 2011 eruption corresponds to $N_{2011}=0$.  With this, we have $P_{pre}=0.07622950\pm0.00000008$ days. 

\subsection{Extinction $E(B-V)$}

The extinction of T Pyx is required for various attempts to get the distance, luminosity, and accretion rate.  The extinction can be quantified by either the color excess, $E(B-V)$, or the V-band dimming in magnitudes, $A_V$.  These are related as $A_V = R_V E(B-V)$.  For the nearby and normal sightline to T Pyx, we can be very confident in using the canonical value of $R_V=3.1$.  For a given $R_V$, the dimming in magnitudes over ultraviolet/optical/infrared wavelengths is accurately given by the analytic equation of Cardelli, Clayton, \& Mathis (1989), as a function only of the parameter $E(B-V)$.

Various values for the extinction to T Pyx have been reported in the literature.  Bruch, Duerbeck, \& Seitter (1981) used an {\it IUE} spectrum of T Pyx to derive $E(B-V)=0.35\pm0.05$ mag as based on the broad 2175\AA~dip.  A later and better analysis of many more {\it IUE} spectra with the same method yielded $E(B-V)=0.25\pm0.02$ mag (Selvelli et al. 2003; Gilmozzi \& Selvelli 2007).  Weight et al. (1994) quote $E(B-V)=0.08$ mag, but do not mention the method or source of this result.  

Shore et al. (2011) find $E(B-V) \approx 0.5\pm0.1$ as based on the strength of diffuse interstellar bands and some correlations between these strengths with $E(B-V)$.  However, their text gives a different value from their abstract, reporting instead $E(B-V) = 0.49\pm0.19$, with the difference of a factor of two between the error bars apparently being due to the averaging of $E(B-V)$ values from four bands.  (Our detailed repetition of this procedure yields an average extinction of 0.47 mag for T Pyx with their reported equivalent widths and their choice of four bands.)  But this improvement by a factor of 2 by averaging over 4 bands is not valid, because the equivalent widths of the bands are very tightly correlated (Friedman et al. 2011) so they are not independent.  Indeed, the uncertainty should not be given by the fractional RMS scatter of the not-independent values from the four bands.  A good way to calculate the real uncertainty is to repeat the procedure of Shore et al. for the many stars in Friedman et al., and calculate the RMS scatter of the average derived $E(B-V)$ values for those stars with similar $E(B-V)$ as claimed for T Pyx.  With this, we find that the one-sigma error bar is 49\%.  Thus, the data and method of Shore et al. are actually producing $E(B-V)=0.47\pm 0.23$.

Here, we report on a new measure of the extinction to T Pyx.  This is based on the {\it GALEX} spectrum, where the 2175\AA~ dip is prominent.  We have dereddened the fluxed and binned spectrum using the extinction model in Cardelli, Clayton, \& Mathis (1989).  With a chi-square fit, we find the $E(B-V)$ value for which the resultant spectrum from 1440-2844\AA~is best fit by some power law.  For this fit, we have not included the bins with the C IV emission line at 1550\AA~or the He II emission line at 1640\AA.  Our best-fit dereddened spectrum and the best-fit power law model are shown in Figure 6.  With this, we get $E(B-V)=0.239\pm0.004$ mag.  This is essentially the same result from the same method as Selvelli et al. (2003), except that we are independent of them and have used a different satellite for the ultraviolet spectrum.

The standard way to determine the {\it total} extinction along a line of sight is to use the far infrared maps of {\it IRAS} and {\it COBE/DIRBE} as given by Schlegel, Finkbeiner \& Davis (1998).  Their maps show T Pyx to be on the edge of the Milky Way.  Their extinction along the entire line of sight (for an assumed value of $R_V=3.1$) is $E(B-V)=0.275$ mag (with a range of 0.253 to 0.290 mag).  This sets a confident upper limit on the possible interstellar extinction to T Pyx.  So we know that the value reported by Shore et al. is roughly one-sigma high, which is to say about a factor of two high.  This extinction limit is equal to that measured for T Pyx.  This means, not surprisingly, that T Pyx is outside most or all of the interstellar medium around the plane of our galaxy that is causing the extinction.

So what is the extinction to T Pyx?  The only confident and accurate method for determining the extinction to T Pyx is based on the 2175\AA~absorption in the ultraviolet spectrum.  For this, the best measure is that by Selvelli and coworkers.  So we adopt their $E(B-V)=0.25\pm0.02$ mag.

\subsection{Distance $D$}

Distances are always critical in astronomy, and the distance to T Pyx, $D$, will be needed to get the luminosity and accretion rate.  Prior work has not turned up any decisive measure of $D$, although needing some answer, a variety of methods have been applied.  All papers hearken back to the original distance estimate of Catchpole (1969), who used three different means, involving the strength of the interstellar calcium absorption line, the velocity of the interstellar calcium line, and the rate of decline in brightness.  Additional methods key off the ultraviolet brightness in comparison with some model, as well as observations of the nova shell.  All of these methods have variations, such as to the lines and the data source used.  With various combinations of these methods and differing input data, many authors have come up with distances from 1000 pc up to $\geq$4500 pc (Catchpole 1969; Webbink et al. 1987; Patterson et al. 1998; Selvelli et al. 2008; Sion 2010; Schaefer 2010; Shore et al. 2011).  Unfortunately, with recent results and new analysis, all these distance determinations are seen to either be wrong or with such large uncertainties as to make them useless.

Catchpole (1969) started with the method that gets the distance from the equivalent widths of the interstellar calcium K line and the interstellar sodium D lines, $EW(Ca)$ and $EW(Na)$ both measured in kilometers per second.  He used a calibration from Beals \& Oke (1953) that $D=34.83\times EW(Ca)$ and $D=30.75\times EW(Na)$, with the distance in parsecs.  Unfortunately, this method, and all similar analyses giving distances based on interstellar absorption, suffer from one large problem and one error , either of which rob the method of the ability to give any useful distance.  The large {\it problem} is that the calibration of many sight lines shows a wild variation in the ratio of distance to absorption.  The one sigma scatter in the distance for some observed equivalent width is half an order of magnitude (e.g., Hunter et al. 2006), so the one-sigma range in derived distance will be one full order of magnitude in distance.  That is, some heavily averaged relation might be linear, but any one star can vary around this average by large factors, with the implication that any absorption distance to T Pyx has an uncertainty of around an order of magnitude.  The {\it error} arises because the linear relation between absorption and distance requires that the entire line of sight lie close to the galactic plane.  The calibration of these relations uses sources almost all of whose whole lines of sight lie within 100 parsecs of the plane (Beals \& Oke 1953; Munch 1968; Hunter et al. 2006; Friedman et al. 2011).  The scale height of interstellar absorbers is around 150 pc, so any sightline extending outside this height above the galactic plane will be in error.  T Pyx is at a galactic latitude of +9.7$\degr$, so its sightline will rise 150 pc above the plane for a distance of 900 pc.  For all $D>2600$ parsec, the sightline will have passed outside three scale heights and have broken into regions that are free of interstellar absorption.  That is, the equivalent widths in interstellar lines will be identical for distances of 2600 pc, 3500 pc, 4500 pc, or 100,000 pc.  Indeed, given the thin nature of the interstellar medium more than $\sim$200 pc above the plane as well as the unknowable natural variations along any line of sight, the absorption at 1200 pc is indistinguishable with that at 100,000 pc.  The procedural error (assuming a linear distance-absorption relation for a distant source at a galactic latitude of +9.7$\degr$) invalidates all published distances involving the strength of interstellar absorption.  It is disconcerting that this mistake was made in the first place, then repeated many times over the last 44 years, and was never caught until now.  There is precedent for this debacle in the case of the recurrent nova RS Oph, where Hjellming et al. (1986) made the same mistake (with RS Oph at a galactic latitude of +10.4$\degr$) so as to get an erroneous distance of 1600 pc, with this distance being uncritically adopted in paper after paper by the bandwagon effect (c.f. Schaefer 2008) so that most papers in the last decade give 1600 pc as the canonical distance by dint of sheer repetition (Schaefer 2009).  For T Pyx, the bottom line is that all distances involving the strength of interstellar absorption are now know to be invalid.

Catchpole (1969) also was the first to try to use the velocity structure of the interstellar absorption lines to get the distance to T Pyx.  Alas, again, this distance determination, and all similar determinations, are fatally flawed.  Both Catchpole (1969) and Selvelli et al. (2008) noted that the galactic coordinates of T Pyx (longitude 257.2$\degr$ and latitude +9.7$\degr$) are near to a node of the rotation curve, so both declined to use the method because the uncertainty will inevitably be large. Shore et al. (2011) note that the interstellar absorption line for CH has a radial velocity of +49.5 km/s and derived a minimum distance of 4,000 pc.  Unfortunately, they had made an error in velocity requiring a correction of -19.6 km/s, so in a published corrigendum, they give the CH line with a real radial velocity of +29.9 km/s as the highest velocity absorption feature in the T Pyx spectrum.  No analysis is given as to the updated distance estimate in their published corrigendum, although Shore et al. (2013) have modified the claimed distance from their first paper from the original $\geq$4500 pc to $\geq$3500 pc.  These limits were obtained by taking this radial velocity and formally applying a galactic rotation curve.  Unfortunately, such a procedure does not account for normal variations in cloud velocities from the galactic average.  In particular, nearby stars in the same direction have clouds absorbing up to +50 km/s.  To take three nearby examples (Hunter et al. 2006), HD 72067 (galactic longitude 262.1$\degr$ and latitude -3.1$\degr$) is at a distance of 488 pc and has an intervening cloud at +30 km/s, HD 74966 (galactic longitude 258.1$\degr$ and latitude +3.9$\degr$) is at a distance of 658 pc and has an intervening cloud at +50 km/s, while HD 68761 (galactic longitude 254.4$\degr$ and latitude -1.6$\degr$) is at a distance of 1473 pc and has an intervening cloud at +50 km/s.  So, T Pyx, with an intervening cloud at +29.9 km/s, need only be further away than roughly 500 pc.  That is, this whole method can only produce a meaningless lower limit.

Many papers have used the method called the `maximum magnitude - rate of decline' (MMRD) relation, to go from the speed class of the light curve (quantified by $t_3$ or $t_2$) to get the peak absolute magnitude, and then the distance.  Unfortunately, the observed scatter in the MMRD for novae in our Milky Way has a total range of 3 mag for decline rates near that of T Pyx (Downes \& Duerbeck 2000), and this translates into a distance uncertainty of a factor of 4.  So even if we take the MMRD at face value, we are left with too large an uncertainty to be useful.  However, there is substantial evidence that the MMRD cannot successfully be applied to RNe, in particular, three of the ten galactic RNe (V2487 Oph, U Sco, and V394 CrA) have MMRD distances far outside our galaxy, while two RNe (U Sco and RS Oph) have MMRD distances in large disagreement with the reliable blackbody distances to the companion stars (Schaefer 2010).  So the MMRD is not confidently applicable to RNe, and T Pyx is such an weird and unique system that the physical conditions of the eruption are likely to be so unusual that the MMRD does not apply.  To make matters much worse, Kasliwal et al. (2011) have found that the novae in M31 show only a very wide scatter with no hint of an MMRD in any form.  Recently, this same result has been found for another galaxy.  With this, the very existence of the MMRD is disproven.  There is no MMRD for nova sets with known distances, while our Milky Way showed an MMRD, so we think that this implies that distances to novae in our Milky Way are subject to substantial band-wagon effects.  In summary, all MMRD distances have a uselessly large real uncertainty and the best idea is that the MMRD neither applies to T Pyx nor any other nova.

A similar distance method, with a physical basis instead of an empirical basis for the absolute magnitude at peak, is to presume that the nova cannot get much above the Eddington luminosity.  Selvelli et al. (2008) calculates that novae with high mass white dwarfs have an Eddington luminosity that corresponds to an absolute magnitude of -6.78.  With this assumption, we can indeed derive a distance.  Unfortunately, we know that the assumption of the Eddington luminosity as a limit is greatly wrong, even for slow novae.  Kasliwal et al. (2011) show that 80\% of novae in the Andromeda galaxy violate the Eddington limit by factors from 2 to 16.

Both Selvelli et al. (2008) and Sion et al. (2010) have derived distances to T Pyx based on the de-reddened ultraviolet flux as compared to the models of Wade \& Hubeny (1998).  There are many large technical problems with this program.  First, the model of Wade \& Hubeny only goes up to white dwarf masses of 1.21 M$_{\odot}$ and up to $10^{-8.5}$ M$_{\odot}$ yr$^{-1}$, thus requiring far extrapolations.  Second, the best fit model has a greatly different power law slope than is observed, and this shows that the model is not applicable to T Pyx.  Third, the ultraviolet luminosity has a strong dependence on the accretion rate, which might only be known to within a factor of two or so, so any derived distance is merely returning the worker's assumption on the accretion rate.  Fourth, the distance derived by Sion et al. (2010) is 1000 pc, while the distance derived by Selvelli et al. (2008) is $\sim$3150 pc.  That both were made with essentially identical data and technique demonstrates that the real accuracy of the method is a factor of three or worse.  Fifth, the model presumption is that the ultraviolet luminosity is provided by a simple accretion disk, but this presumption is violated by T Pyx with the case that a magnetic field from the white dwarf is disrupting any disk (Schaefer \& Collazzi 2010; Oksanen \& Schaefer 2011) and with the case that there is continuing nuclear burning on the white dwarf (Patterson et al. 1998).  More generally, and more critically, we prove that the T Pyx light is not coming from an accretion disk (see Sections 5.2.2 and 5.2.3).  All this is to say that any model with a simple accretion disk is invalid and its application will give a wrong distance.  In all, the method of comparing the ultraviolet flux to models results in uncertainties that are too large to produce any useful distance estimate.

T Pyx has a nice shell, so might it be possible to get a distance from this?  In particular, the expansion parallax distance method is the traditional means to get nova distances thought to be good.  But this method is beset by several factor-of-two uncertainties that combine to make for poor real error bars, with these problems always being ignored.  (1) It is unclear what angular radius to assign to the shell in any image, partly because all nova shells are out of round, often significantly, partly because many novae (including T Pyx) have multiple shells, and partly because it is unknown as to which isophotal level (e.g., the outermost, the half-peak, the peak) should be used.  (2) It is unclear as to what velocity from a line profile to use, with possibilities including the HWHM and the HWZI and differing by a factor of two.  (3) Nova shells are generally far from spherical in shape, often being bipolar, such that the expansion velocity along the line of sight can easily be up to a factor of two different from a tangential expansion velocity.  These problems show that expansion parallaxes have a real uncertainty of at least a factor of two in distance.  For the case of T Pyx, we cannot use the expansion parallax for two reasons.  First, the explosion is likely bipolar (Shore et al. 2013; Chesneau et al. 2011), so the relation between the transverse expansion velocity and the radial velocity can only be speculative.  Second, the observed shell actually comes from a slow-velocity classical nova eruption around the year 1866 (Schaefer, Pagnotta, \& Shara 2010a), so we have no spectrum to get any radial velocity.  (The later RN eruptions have completely different expansion velocities, but we see no shell from them.)  So expansion parallax cannot yield any distance, no matter how crude, to T Pyx.

T Pyx has a pre-existing shell, so in principle, we can watch the light echo of an eruption sweeping through the clumps in the shell so as to get a distance.  The idea would be to measure the polarization of the echo light as a function of angular radius, where the peak in the polarization fraction corresponds to light that is single scattered at a right angle.  With the geometry of the nova, we can watch the ring of maximal polarization expand at the speed of light, with the angular size equaling the light travel distance since peak light.  Unfortunately, the 2011 eruption of T Pyx had the ring of maximum polarization sweeping through the shell around the time of solar conjunction, and no polarization data were taken.

The preceding seven paragraphs present a horrifying case where all prior distance estimates to T Pyx are either erroneous, not applicable, and/or uselessly poor.  (Indeed, similar analysis can be made for most novae distances, with the only reliable distances being for the few novae with parallaxes from the {\it Hubble Space Telescope}, novae in external galaxies, novae in galactic clusters, and novae with blackbody distances to the companion star.)  But all is not lost, as there are three constraints that can be confidently placed.  (1) The first constraint is that the interstellar absorption is significant and comparable with the total galactic extinction along the line of sight.  With this we can be confident that T Pyx is at a distance such that its line of sight passes through most of the ISM in the disk, and it must rise more than something like 150 pc above the galactic plane.  With suitably round numbers, we know that D is greater than $\sim$1000 pc.  (2) There is some real upper limit on the peak absolute magnitude of novae, and this is far above the Eddington limit.  The M31 novae have the most luminous event at absolute magnitude -10.7 (Kasliwal et al. 2011), and T Pyx will confidently not exceed this.  The theoretical models of Yaron et al. (2005) come with the same peak.  For $A_V=0.75$ mag and a V-band peak of 6.4 mag, the upper limit on D is 37,000 pc, which is far outside our galaxy.  (3) T Pyx is in the Milky Way and is very unlikely to be in the sparse outer regions.  A distance of 10,000 parsecs would place T Pyx at a height of 1,700 pc above the plane and at a galactocentric distance of 14,000 pc.  Somewhere around this position will have a sufficiently low star density that no one would care to say that T Pyx is farther.  And this is all we have that we can have any confidence in.  Thus, we conclude that T Pyx has a highly uncertain distance with only poor constraints, such that $1000 \lesssim D \lesssim 10,000$ parsecs.

\subsection{Stellar Masses $M_{WD}$ and $M_{comp}$}

Theory has a very strong result that RNe must have the mass of the white dwarf ($M_{WD}$) close to the Chandrasekhar mass, roughly $1.2<M_{WD}<1.4$ M$_{\odot}$ (e.g., Starrfield, Sparks, \& Shaviv 1988; Prialnik \& Kovetz 1995; Hachisu \& Kato 2001).  The reason is that only high mass white dwarf will have a high enough of a surface gravity to compress a thin layer of accreted material to the kindling temperature for hydrogen for an amount of material that can be accreted in under one century.  Observational tests from radial velocity curves of this theoretical claim have only turned up patchy agreement.  Published $M_{WD}$ values have been $>2.1$ M$_{\odot}$ for T CrB (Kraft 1958), 0.29$\pm$0.14 M$_{\odot}$ for U Sco (Johnston \& Kulkarni 1992),  1.55$\pm$0.24 M$_{\odot}$ for U Sco (Thoroughgood et al. 2001),  1.37$\pm$0.13 M$_{\odot}$ for T CrB (Stanishev et al. 2004), and 0.7$\pm$0.2 M$_{\odot}$ for T Pyx (Uthas et al. 2010).  This dismal record for radial velocity analyses of RNe is matched by their dismal record for getting $M_{WD}$ in all cataclysmic variables, with the reason being that the the emission lines come from the inner disk and do not perfectly follow the dynamics of the white dwarf.

Both Selvelli et al. (2008) and Schaefer et al. (2010a) have used information on the recurrence time scale, the accretion rate, and the ejection velocity (for both the RN events and the 1866 classical nova eruption) so as to try to narrow the range of $1.2<M_{WD}<1.4$ M$_{\odot}$.  The typical analysis is to interpolate within the grid of theoretical models in Yaron et al. (2005) so as to find the white dwarf mass and accretion rate that provides the closest match in recurrence time scale and ejection mass.  Schaefer et al. (2010a) came to the conclusion that $1.25<M_{WD}<1.30$ M$_{\odot}$.  Selvelli et al. (2008) finds the closest agreement with the models has $M_{WD}\approx 1.33$ M$_{\odot}$, with an uncertainty $\sim$0.03 M$_{\odot}$.  With no better information, we will here take the middle ground that is acceptable to both studies and conclude that $M_{WD}=1.30\pm0.05$ M$_{\odot}$.  

Into this situation, Uthas et al. (2010) report a radial velocity curve for T Pyx, with the results that the companion star mass, $M_{comp}$, is 0.14$\pm$0.03 M$_{\odot}$, $M_{WD}$ is 0.7$\pm$0.2 M$_{\odot}$, the mass ratio, $q=M_{comp}/M_{WD}$, is 0.20$\pm$0.03, and the orbital inclination is 10$\degr \pm$2$\degr$.  Here, the given white dwarf mass is in large contradiction to the very strong theory.  Fortunately, it turns out that their given $M_{WD}$ can easily be changed, as they allow in their paper.  Without accurate inclination measures (say, from an eclipse or ellipsoidal effects), we need to only slightly change the inclination and $M_{WD}$ would become a lot larger.  What Uthas et al. are actually measuring are the amplitude of the radial velocity curve for the emission lines plus the wavelength separation between peaks in the emission line profiles.  From these two measures, they use a non-standard (but reasonable) technique to derive the mass ratio.  Their main line of argument is then to claim that the companion should follow a slightly inflated main sequence mass-radius relation (with the radius known from the Roche lobe geometry) so as to come up with a companion mass of 0.14$\pm$0.03 M$_{\odot}$, and then use the mass ratio to get $M_{WD}$.  But it is a large leap of faith to expect that such a weird system as T Pyx will have a normal main sequence companion.  So we are left with a dilemma, either that the white dwarf is half of the Chandrasekhar mass in contradiction to strong theory or that the companion star is not a normal main sequence star.  Faced with this dilemma, we have to choose that the companion star is not a normal main sequence star.  With this, we have a near-Chandrasekhar mass white dwarf and with their mass ratio, we get $M_{comp}=0.26\pm0.04$ M$_{\odot}$. 

Another possibility is that the mass ratio is the problem, specifically with the emission line profile {\it not} coming from a disk.  In Section 5.2.2, we prove that the {\it continuum} light can at most have only a small fraction coming from an accretion disk, so it is reasonable to think that the emission lines do not come from a disk.  (There are other possible explanations for the double peaked profile, including from the pair of accretion columns.)  In the absence of a disk, the argument to go from the separation of the peaks to the mass ratio is broken.  Then, we still have $M_{WD}=1.30\pm0.05$ M$_{\odot}$, but now we can accept the argument based on the orbital period (plus some assumed sequence of stellar structures) that $M_{comp}=0.14\pm0.03$ M$_{\odot}$.  With this, we would have $q=0.108\pm0.023$.  

We do not have a clear answer as to whether the companion has a mass of 0.14 or 0.26 M$_{\odot}$.  The problem is that there are three claims (the white dwarf is near the Chandrasekhar mass, the mass ratio comes from the peak separation in the line profiles, and the companion satisfies some mass-radius relation) that are contradictory.  Various researchers will pick out their favorite claim and derive their conclusion.  For example, we know that RNe require near-Chandrasekhar mass white dwarfs.  With all three claims denied in turn, the companion is still either 0.14 or 0.26 M$_{\odot}$.

\subsection{Accretion Rate $\dot{M}$}

We can use our data and analysis to address the question of the quiescent accretion rate.  The value of $\dot{M}$ is key for understanding a variety of peculiarities in the T Pyx system.  The most critical need for $\dot{M}$ is so that we can calculate how much mass has arrived on the white dwarf from 1967 to 2011, for a comparison with the mass ejected by the 2011 eruption.  This comparison will then decide whether the white dwarf has gained or lost mass over the entire eruption cycle, and hence whether it can be a Type Ia supernova progenitor.  Here are five methods for determining $\dot{M}$, some new, some old, and some failing.

Method 1:  The orbital period of T Pyx has been changing closely linearly with time (see the good parabolic shape of the O-C curve in Figure 8).  This large and steady period change can only arise from the mass transfer in the system.  The observed increase in the orbital period is expected when a low mass donor star transfers matter to the high mass white dwarf.  The period change will be proportional to $\dot{M}$, so we can use the observed period change to deduce the accretion rate.  The default scenario is conservative mass transfer, in which all the mass leaving the donor star arrives onto the white dwarf, and this material carries with it the specific angular momentum of the donor star.  For conservative mass transfer, 
\begin{equation}
\dot{M} = (M_{comp}/[3-3q])/(P/\dot{P}).
\end{equation}
With $M_{comp}=0.26$ M$_{\odot}$, we have $\dot{M}=3.5\times10^{-7}$ M$_{\odot}$ yr$^{-1}$.  With $M_{comp}=0.14$ M$_{\odot}$, we have $\dot{M}=1.7\times10^{-7}$ M$_{\odot}$ yr$^{-1}$.  The conservative case is where all matter leaving the companion falls onto the white dwarf, with no angular momentum lost to the system.  This case has to be fairly close (but certainly not exact) for T Pyx.  Depending on the details of how the accretion process moves mass and angular momentum, the accretion rate can vary somewhat from this derived $\dot{M}$.  Within this moderate uncertainty, this basic result is robust, depending only on timing measures and on simple dynamics of the binary.  We will express the allowed range of the accretion rate as $\dot{M}=1.7-3.5\times10^{-7}$ M$_{\odot}$ yr$^{-1}$.

There are alternative possibilities to produce a good parabola in the O-C curve, so we should consider these cases.  Here are the three other processes, all of which fail to explain T Pyx:  (1) The O-C curves for cataclysmic variables occasionally show a sinusoidal oscillation with periods of years, likely associated with activity cycles on the companion star (Bianchini 1990).  If only a portion of the sine wave O-C curve is observed, the curve might appear as a parabola, for example, when Robinson et al. (1995) found that the Z Cam had a nearly parabolic O-C curve from 1972-1992, yet when the O-C curve was extended to 2002 the shape looked more like one period of a sinewave (Baptista et al. 2002).  We have fit our O-C curve for T Pyx to a sine wave, and find acceptable fits only for periods longer than 80 years and half-amplitudes larger than 0.2 days.  This is greatly different than all solar cycle variations in cataclysmic variables (Baptista et al. 2003), because these all have periods less than 30 years (with the median value near 10 years) and O-C half amplitudes of less than 3 {\it minutes} (with the median value 1.2 minutes).  The cataclysmic variables below the period gap have a maximum half-amplitude of less than one minute (0.0007 day), and so this effect is a factor of $>300\times$ too small to account for T Pyx.  (2) The O-C curve will also change as a sinewave for the case of T Pyx being in far orbit around a third star, with the visible parabola merely being a part of the sinewave.  For such a possibility, the orbital period of the third body would have to be $>80$ years and the orbital radius of the nova binary system would be $>0.2$ light-days, or $>35$ AU.  From Kepler's Law, any third body must currently be $>9$ M$_\odot$ and its progenitor would have exploded as a supernova before the white dwarf could have formed in the system.  Even if such a triple could have remained bound, we do not think that anyone will try to claim that T Pyx has a massive black hole as a very-wide third body.  (3) Cataclysmic variables {\it above} the period gap suffer a steady period change due to magnetic breaking (Patterson 1984), and this will lead to a parabolic O-C curve.  Importantly, this magnetic breaking turns off {\it below} the period gap, so this process would not be applicable to T Pyx.  Nevertheless, suspicions have been raised of some small residual magnetic breaking below the gap, so we should still consider this possibility.  This possibility is strongly refuted by either of two means.  First, the magnetic breaking mechanism as applied to long period cataclysmic variables always produces O-C changes with $P/\dot{P}>5,000,000$ years (Patterson 1984), so a small residual effect for short period systems must have $P/\dot{P}\gg10^{8}$ years or so, whereas T Pyx has $P/\dot{P}=313,000$ years.  Second, the magnetic breaking robs the system of angular momentum, so the orbital period would be {\it decreasing}, whereas T Pyx has its orbital period {\it increasing}.  In summary, the three alternative possibilities all completely fail.

The parabolic O-C curve is caused by the normal mass transfer.  We know this because all alternatives fail completely.  We also know this because the behavior seen (period increasing, $P/\dot{P}=313,000$ years, and a good parabola shape) is just that expected from strong theory (see the discussion for the next two methods).  We also know that the parabolic O-C curve comes from mass transfer because the small deviation from 2008-2011 corresponds with the drop in the accretion rate.  The measured period change comes from timing observations (always good for high reliability) and the derivation of $\dot{M}$ is from a purely dynamical method (where the physics is simple and well-known, being independent of distance, extinction, and anyone's theory).  The derive accretion rate has some uncertainty associated with the not-perfectly known companion mass, and some additional comparable uncertainty will arise from the mass transfer not being perfectly conservative.  But within these error bars, our derived $\dot{M}$ is robust and by far the best measure for T Pyx.  In summary, we find $\dot{M}=1.7-3.5\times10^{-7}$ M$_{\odot}$ yr$^{-1}$.

Method 2:  We can set a very useful upper limit on the accretion rate onto the white dwarf, because if $\dot{M}$ is too high then the hydrogen on the surface of the white dwarf will undergo steady nuclear burning.  We see a normal nova eruption, so we know that there is no steady hydrogen burning, so the accretion rate must be below some limit.  The limit for steady hydrogen burning is $2-3\times10^{-7}$ M$_{\odot}$ yr$^{-1}$ for a near Chandrasekhar mass white dwarf (Nomoto et al. 2007; Shen \& Bildsten 2007).  A substantial problem for this method is that the quoted steady hydrogen burning limit is for a white dwarf that is in thermal equilibrium with the given accretion rate, whereas T Pyx has a complex history with rates apparently going from $10^{-11}$ to $10^{-6}$ M$_{\odot}$ yr$^{-1}$ in the recent past.  With a cooler white dwarf than expected for the current accretion rate, the threshold for steady hydrogen burning should be raised by some unknown amount.  With this, the accretion rate limit from this method might have to be somewhat raised.

Method 3:  The 2011 eruption ties the inter-eruption interval to 44 years.  With the approximate mass of the white dwarf, we can estimate the required trigger mass on the white dwarf , so the accretion rate will be just the trigger mass divided by the inter-eruption interval.  A substantial problem is that the accumulated mass required for the trigger has a strong dependence on the adopted conditions.  It varies by over an order of magnitude over the acceptable range of white dwarf masses (Yaron et al. 2005), it varies by a factor of four over a plausible range of metallicities, and it varies by a factor of five over a plausible range of $^3$He mass fraction (Shen \& Bildsten (2007).  In addition, the trigger mass varies by a factor of two with a change in $\dot{M}$ by a factor of ten over the plausible range.    Nevertheless, for the estimated conditions, the T Pyx white dwarf must have accumulated $ 10^{-6.0\pm 0.3}$ M$_{\odot}$ (Yaron et al. 2005), $ 10^{-6.3\pm 0.2}$ M$_{\odot}$ (Nomoto et al. 2007) or $10^{-5.5\pm 0.3}$ M$_{\odot}$ (Shen \& Bildsten (2007) to be able to erupt.  This range of theoretical models can be represented as $10^{-5.9\pm 0.6}$ M$_{\odot}$.  With the 44 year interval, we get the average rate of deposition of mass onto the white dwarf is in the range $10^{-7.5\pm0.6}$ M$_{\odot}$ yr$^{-1}$.  Importantly, this highly uncertain estimate is based on specific theoretical models, so we must be careful to not use this estimated $\dot{M}$ to evaluate future models.

Method 4:  We can compare our {\it GALEX} spectrum with the accretion disk models of Wade \& Hubeny (1998) and pick out the accretion rate that best reproduces the model.  The available models have the two highest mass white dwarfs of 1.21 and 1.03 M$_{\odot}$, while the highest mass model has the two highest accretion rates of $10^{-8.5}$ and $10^{-9.0}$ M$_{\odot}$ yr$^{-1}$, so we have to extrapolate far outside the range of available models to get to the conditions likely to be relevant for T Pyx.  The models have wavelength range of 850-2000\AA, while the binned {\it GALEX} spectrum has a range of 1456-2828\AA, so the overlap for comparison is fairly small from 1456-2000\AA.  Over this range, the observed spectrum is featureless (with the two emission lines not being relevant for the disk models) so all we can get is a spectral slope.  Over this same range, for low inclination systems, the models all are largely featureless (certainly for features large enough to be visible in the {\it GALEX} spectrum) and have essentially the same spectral slope.  In this situation, there is no way for the {\it GALEX} spectrum to distinguish the disk model with the best accretion rate.  This same analysis and conclusion also holds for the {\it IUE} spectrum.  In all, this hopeful method does not work in practice.  And in principle, the likelihood that T Pyx is a magnetic system (with no disk or a truncated disk) and/or has non-disk light sources dominating in the ultraviolet implies that the disk models are not applicable anyway.

Method 5:  The accretion luminosity dominates the light from T Pyx, and this is proportional to the accretion rate, so in principle we can take the observed flux from the system and derive $\dot{M}$.  Typically, this will involve making some estimate of the unobserved flux in the far ultraviolet, then performing an integral over the spectral energy distribution (as in Section 3.4), and finally multiplying by $4\pi D^2$ for some assumed distance.  Selvelli et al. (2003) report such a calculation based on their {\it IUE} spectrum, deriving accretion rates from $2.2-4.6 \times 10^{-8}$ M$_{\odot}$ yr$^{-1}$.  Selvelli et al. (2008) used a similar method to determine the accretion rates as $1.1 \pm 0.3 \times 10^{-8}$ M$_{\odot}$ yr$^{-1}$, while they also report on various relations between $\dot{M}$ and the V-band absolute magnitude to come up with similar accretion rates.  Unfortunately, all such calculations have real uncertainties of over four orders of magnitude in $\dot{M}$.  One reason is that the extrapolation to the unobserved ultraviolet leads to a range of a factor of $10^4 \times$ depending on where the break in the the spectral energy distribution is in the range from $2\times10^{15}$ Hz to $2\times10^{17}$ Hz (see Section 3.4).  And there is a another independent factor of 100$ \times$ uncertainty simply due to the distance to T Pyx being uncertain by a factor of ten.  So, for acceptable input, we could get a luminosity as low as $2 \times 10^{34}$ erg s$^{-1}$ and as high as $2 \times 10^{40}$ erg s$^{-1}$.  Any $\dot{M}$ derived in this way will have a real uncertainty of six orders of magnitude, and that makes this method useless.

Method 6:  T Pyx is at best only a faint X-ray source, and this can be used to place limits on the accretion rate.  Selvelli et al. (2008) report that their observations with {\it XMM-Newton} for 22.1 ksec in 2006 shows a faint source visible over the 0.2-8 keV energy range.  The signal was too faint to allow for a reliable fit, but they demonstrate that a 200,000$\degr$K blackbody with $10^{37}$ erg s$^{-1}$ source at 3.5 kpc is rejected because the spectrum below 0.5 keV would be roughly 300 times brighter than observed.  For the soft X-ray component, this would imply an X-ray luminosity of roughly $10^{34.5}$ erg s$^{-1}$.  Balman (2010) used the same {\it XMM-Newton} data set from 2006 (except she reports on effective exposure times of 30.6, 39.7, and 38.8 ksec for the three detectors).  She found that an X-ray image shows an extended profile, with the unstated fraction of light above a single point source certainly coming from the shell around T Pyx.  A spectral fit gives a luminosity of $6.0 \times 10^{32}$ erg s$^{-1}$ at a distance of 3.5 kpc over 0.3-10.0 keV.  If this X-ray luminosity is taken to be the light from the accretion onto the white dwarf ($\frac{1}{2} G M_{WD} \dot{M} / R_{WD}$), then the accretion rate is $8 \times 10^{-11}$ M$_{\odot}$ yr$^{-1}$.  If the X-ray luminosity is taken to be from steady hydrogen burning of the freshly accreted material on the surface of the white dwarf ($0.007X\dot{M}c^2$ if all the hydrogen is burned, where $X$ is the usual hydrogen fraction), then the accretion rate is around $2 \times 10^{-12}$ M$_{\odot}$ yr$^{-1}$.  But this statement is made with the assumption that all of the accretion luminosity is coming out in the X-ray, whereas this is clearly not true.  Most of the accretion energy is coming out in the ultraviolet or the far ultraviolet.  So this method really is only producing a {\it limit}, with $\dot{M} \gg 2 \times 10^{-12}$ M$_{\odot}$ yr$^{-1}$.

We have given a full discussion for all six methods because they have already appeared in the literature.  We have found methods 4-6 as being useless or inapplicable.  Methods 2 and 3 are based on theoretical models, which have to be right, although the real uncertainties are larger than we would hope for.  Only method 1 returns a reliable and robust measure of $\dot{M}$, and even this has an uncertainty of roughly a factor of two.  So, finally, we conclude that $\dot{M}=1.7-3.5\times10^{-7}$ M$_{\odot}$ yr$^{-1}$.

\section{Analysis}

In the following subsections, we address a variety of questions and topics related to a model for T Pyx:

\subsection{The Rise to Peak}

T Pyx has a well-defined light curve with essentially continuous coverage and BVI colors starting at the the time of the discovery (see Figures 2 and 3, and Table 2).  The nova rose from V=13.0 (discovery) to V=10.0 in a time of 8 hours, for an average rate of 9.0 mag day$^{-1}$.  This fast rise slows within the first day, rising to V=7.9, 2.0 days after discovery.  Then, T Pyx fades by a third of a magnitude over the next day, followed by a very slow rise back to V=7.9 over the next five days.  Soon after this, the rise increases to a minor peak of V=6.7 on JD2455685 (19 days after discovery) and then up to the highest peak of V=6.4 on JD2455693 (27 days after discovery).

The initial rise of a nova can be simply modeled by a uniformly expanding shell.  The observed flux from the system will equal
\begin{equation}
F = F_0 + (v \tau /D)^2 F_{\nu}(T).
\end{equation}
Here, $F_0$ is some constant flux from the disk and companion star, $v$ is the expansion velocity of the outer edge of the shell (corresponding to the effective photosphere), $\tau$ is the time of expansion, $D$ is the distance to the T Pyx system, $T$ is the effective temperature of the photosphere, and $F_{\nu}(T)$ is the flux from a unit area of the photosphere.  For the case of bolometric flux, $F_{\nu}(T)=\sigma T^4$ for a blackbody photosphere.  Equation 5 can be converted into AB magnitudes as 
\begin{equation}
m = -48.60 - 2.5 \log _{10}[F_0 + C(t-t_0)],
\end{equation}
with
\begin{equation}
C = (v/D)^2 F_{\nu}(T).
\end{equation}
Here, $\tau$ is expressed in the more practical form of the time ($t$) minus some fitted zero time for the start of the rise ($t_0$).  $C$ is a combination of values that presumably remain nearly constant throughout the fast initial rise.  Before $t_0$, the magnitude should be a constant corresponding to the flux $F_0$ alone.  We now have a simple model with three unknowns ($F_0$, $C$, and $t_0$) that should describe the initial rise.

We have fitted our T Pyx light curve from the time of discovery until JD2455668.0 with a chi-square analysis.  This includes 2151 magnitudes.  To get a reduced chi-square of unity, we had to adopt a systematic uncertainty of 0.25 mag, which is a measure of systematic errors in the measurements plus deviations from the model.  Our best fit model is shown in Figure 4.  The constant flux corresponds to a magnitude of 14.63, $t_0 = 2455665.967$, and $C=1.56 \times 10^{-23}$ erg s$^{-2}$ cm$^{-2}$ Hz$^{-1}$.  This model fits the initial fast rise well, and gives some confidence that the simple model is close to the real case.  The sudden deviation two days after the initial rise would then indicate some significant departure from the simple model in the sense that the expanding shell appears fainter than expected.  We suggest that this deviation is caused by the outer portion of the ejecta expanding to become optically thin enough so that the photosphere can shrink significantly in size.

We had originally hoped that this rise could be used to determine an accurate distance to T Pyx by the Baade-Wesselink method.  Our light curve fit gives a value for $C$, the spectral line widths will give $V$, and we can use theory or colors to estimate $F_{\nu}(T)$).  Then the distance can be directly derived from Equation 8.  Unfortunately, the inputs are known only poorly.  For the velocity, we do not know what position in a line profile (e.g., half-width-zero-intensity or half-width-half-maximum) corresponds to the velocity at the photosphere.  For $F_{\nu}(T)$, the blackbody approximation is expected to be good for the photosphere, but the early colors are inconsistent with a blackbody.  As such, the temperature cannot be determined with any useful confidence from observation of the colors.  Also, no early spectrum exists, so we cannot get the temperature from spectral lines.  From theory, the temperature varies by a factor of $\sim$3, with a strong dependance on the white dwarf mass and other model details.  With no way to get the temperature to any useful accuracy, we cannot get $F_{\nu}(T)$ even if we know the emission mechanism.  Nor can we turn this around to derive $F_{\nu}(T)$ from $C$, $D$, and $V$, because again the uncertainties of the inputs would be too large to be useful.  In all, even though $C$ is fairly well measured, we cannot get useful information on any of $D$, $V$, or $F_{\nu}(T)$.

\subsection{The Nature of the Emission From Near the White Dwarf}

The secondary star is a small red dwarf with a negligible luminosity, and the white dwarf itself is so small that it cannot provide any significant emission.  So all the observed flux from T Pyx can only come from the accretion luminosity by some means.  This emission can only come from near the white dwarf, perhaps as an accretion disk, an accretion stream, or some more complicated structure.  This subsection gives our analysis as to the nature of this emission.

The mass falling onto the white dwarf has a lot of energy that will be radiated at some wavelength.  We have a fairly good idea of the accretion rate and the white dwarf properties, so we can make a good estimate of the accretion luminosity as $G M_{WD} \dot{M} / R_{WD}$.  With $1.7-3.5\times10^{-7}$ M$_{\odot}$ yr$^{-1}$ falling onto the white dwarf, the luminosity would be $2.7-5.5\times10^{36}$ erg s$^{-1}$.  This luminosity is independent of the large uncertainty in the distance, rather it comes from knowing the accretion rate dynamically from the steady $\dot{P}$.  This accretion luminosity could come out with any of a variety of possible spectral energy distributions, and with some unknown fraction in the unobserved far-ultraviolet. 

\subsubsection{Not a Single Blackbody}

Selvelli et al. (2003) and Gilmozzi \& Selvelli (2007) point out how the {\it IUE} spectrum can be well fit with a single blackbody spectrum of 34,000$\degr$K.  This temperature explains why essentially no X-ray flux is visible with {\it XMM}.  However, as they point out, this blackbody cannot account for the flux outside the ultraviolet regime, because the optical and infrared flux is far above this single blackbody.  This can be seen in Figure 7, where the optical and infrared falls off far from the $f_{\nu} \propto \nu^2$ of the Rayleigh-Jeans tail.  No single blackbody can account for any significant fraction of the T Pyx SED over any but a narrow frequency range.

Patterson et al. (1998) and Knigge et al. (2000) postulated an emission source near the white dwarf that emits a large luminosity of far-ultraviolet and super-soft X-ray photons so as to irradiate the companion star and drive the required high accretion rate in the T Pyx system.  They took this source to be something like a blackbody with a temperature of 240,000$\degr$K and a luminosity of $10^{37}$ erg s$^{-1}$.  Selvelli et al. (2008) have confidently proven that any such source would have been seen with the {\it XMM-Newton} X-ray images in 2006.

\subsubsection{Not an Accretion Disk}

Based on the strong precedent of other novae and cataclysmic variables, the common expectation would be to have the flux coming from some sort of an accretion disk with reprocessing.  Indeed, Gilmozzi \& Selvelli (2007) find that the ultraviolet SED is well fit by $f_{\nu} \propto \nu^{1/3}$, and this is exactly the slope expected for the continuum from disks.  (But this is not a strong argument due to their ultraviolet spectrum also fitting a blackbody.  Further, Frank et al., 2002, point out that for the case of T Pyx with a hot outer edge, the $\nu^{1/3}$ regime ``may be quite short and the spectrum not very different from a blackbody" and ``discs around white dwarfs with $R_{out}\sim 10^2 R_{in}$ [as for T Pyx] do not possess an obvious power law character."  We know that the T Pyx $\nu^1$ spectrum has to turn over, so there must be some region around the ultraviolet where the SED slope will be $\sim \nu^{1/3}$ whether or not there is any disk.)  An additional argument for a disk is that Gilmozzi \& Selvelli (2007) found absorption lines in the ultraviolet that are characteristic of disks.  (This last point is only weak evidence for a disk, because it really only says that a cool gas of ordinary composition is just outside the continuum emission region, and there are many structures that can yield the same absorption lines.  In addition, their detailed comparison with the accretion disk model grids of Wade \& Hubeny, 1998, showed no satisfactory match.)  So the default idea is that the light from T Pyx is dominated by an accretion disk.

But this default idea cannot be correct because the SED is greatly different from that of any accretion disk.  The general shape of accretion disk continua if a $f_{\nu} \propto \nu^{1/3}$ power law over a region (perhaps quite narrow) corresponding to temperatures within the disk, with a Rayleigh-Jeans cutoff ($f_{\nu} \propto \nu^{1/3}$) to lower frequencies (see Figure 5.2 of Frank et al. 2002 and Wade \& Hubeny 1998).  The ultraviolet portion of the T Pyx spectral energy distribution is consistent with a $\nu^{1/3}$ power law, but the entire UBVRIJHK part of the SED has $f_{\nu} \propto \nu^{1}$.  The UBVRIJHK region is much too large a range to be the roll-over from $\nu^{1/3}$ to $\nu^{2}$, and we see zero curvature in Figure 7.  With this, we know that the UBVRIJHK emission is {\it not} from an accretion disk.

For any disk around T Pyx, even the outer edge would be very hot.  Detailed calculation for an $\alpha$-disk and $\dot{M}=3\times10^{-7}$ M$_{\odot}$ yr$^{-1}$ give a temperature of the outer edge (taken to be at 90\% of the Roche lobe radius) to be $T_{outer}$=20,000$\degr$K, or 42,000$\degr$K if reprocessing of the accretion luminosity from the central source is included.  The turnover to a Rayleigh-Jeans slope in the SED occurs at a frequency of $kT_{outer}/h$, with this being either $2.6\times10^{15}$ Hz or $5.5\times10^{15}$ Hz.  (This temperature of the outer edge of any disk in the T Pyx system is substantially hotter than is common for cataclysmic variables, simply because T Pyx has such a high accretion rate and because the short orbital period requires a a very small disk.)   Thus, the entire observed portion of the ultraviolet should already be in the Rayleigh-Jeans region.  This means that the observed $f_{\nu} \propto \nu^{1/3}$ from {\it IUE} has no causal connection with any accretion disk.

\subsubsection{Not a Superposition of Blackbodies}

Is it possible to superpose a number of blackbody spectra so as to produce the observed $f_{\nu} \propto \nu$ SED?  That is, just as an accretion disk SED is a superposition of blackbody spectra from different temperatures within the accretion disk so as to produce a power law SED, perhaps we can find some other distribution of temperatures over the emitting surface area so that the sum of all the thermal spectra will produce the observed SED.  The surface area of the photosphere of the thermal emission must be entirely contained within the Roche lobe of the white dwarf, because the white dwarf motion is visible in the radial velocity curve (not hidden inside some enveloping photosphere) and because any larger structure would be rapidly destroyed by churning from the companion star.  The radius of the white dwarf Roche lobe is close to 0.5 R$_{\odot}$.

Can the near infrared flux come from any superposition of blackbody spectra?  The most efficient case (while avoiding a Rayleigh-Jeans SED) is to have enough surface area with a temperature such that the peak of $f_{\nu}$, at a frequency of $\nu_{max}$, is in the near infrared.  For the K-band flux, we have $\nu_{max}=1.36\times 10^{14}$ Hz and $f_{\nu}=0.0012$ Jy.  The $\nu_{max}$ value corresponds to a temperature of 2300$\degr$K, for which the blackbody equation gives a flux of $F_{\nu}=7.42\times10^{-6}$ erg cm$^{-2}$ s$^{-1}$ Hz$^{-1}$.  These quantities are connected as $4\pi D^2 f_{\nu} = 4\pi R_{sphere}^2 F_{\nu}$, where $R_{sphere}$ is the radius of a sphere that has the required surface area.  The distance $D$ is unknown with only poor limits of $1000\lesssim D \lesssim 10000$ parsecs, even though T Pyx has a `traditional' value of 3500 pc or so.  For the minimal distance of 1000 pc, the K-band flux can only be produced by a region whose surface area equals that of a sphere 1.80 R$_{\odot}$ in size.  The inclusion of any hotter surface area would add flux to the K-band only with this hotter blackbody imposing a Rayleigh-Jeans slope on the SED, and such is not observed.  This required surface area, so as to produce the K-band flux, is greatly larger than that of the Roche lobe, so there is no possibility that any significant amount of the flux comes from any superposition of thermal spectra.  This same conclusion holds true for the R-, I-, J-, H-, and K-band fluxes.  For the `traditional' distance of 3500 pc, the flux with wavelengths longer than 2000\AA ~cannot come from a superposition of blackbodies.  Thus, the red and near infrared flux from T Pyx certainly cannot come from any superposition of thermal spectra, and hence there must be a bright nonthermal source in the system.

We can place a distance-independent limit on the presence of any superposition of blackbodies by realizing that the accretion luminosity (with $\dot{M}=1.7-3.5\times10^{-7}$ M$_{\odot}$ yr$^{-1}$) must come out in the SED.  This accretion rate gives a total accretion luminosity of $2.7-5.5\times 10^{36}$ erg s$^{-1}$.  The most efficient way to get this luminosity out with a superposition of blackbodies is to put all of the surface area to the highest temperature.  (This would immediately violate the observed SED shape, but such a calculation is good to show an extreme limit.)  Gilmozzi \& Selvelli (2007) fit the ultraviolet SED to a temperature of 34,000$\degr$K, and the temperature cannot be much higher without violating the X-ray limit.  A 34,000$\degr$K blackbody emitting $5\times 10^{36}$ erg s$^{-1}$ has a surface area corresponding to a sphere with a radius 1.0 R$_{\odot}$.  That is, even pushed past what is possible, the accretion luminosity requires a surface area that is much larger than that of the white dwarf Roche lobe.  With most of the luminosity coming out in the ultraviolet, we see that the accretion luminosity is too large to come out as any superposition of blackbodies.

\subsubsection{Nonthermal Emission}

With all possibilities of thermal emission strongly rejected, we can only conclude that the T Pyx SED is primarily from non thermal emission.  Nonthermal emission, say from free-free or cyclotron processes, can easily produce a high luminosity within the Roche lobe.  Nonthermal emission can produce $f_{\nu} \propto \nu$ over a wide range of frequency.  So we know that the light from T Pyx is nonthermal.

Nonthermal emission must come from a region that is not optically thick, because if it were optically thick then the photons would rapidly thermalize, form a photosphere and be sent into space with a blackbody spectrum.  This means that we are seeing a {\it volume} of emission, not a {\it surface} of emission.  Likely, any such volume will have increased density near the white dwarf, and some central core might have high enough optical thickness so as to appear as a blackbody, with this thermal light being seen simultaneously with the nonthermal emission from the surrounding volume.  It is unclear what fraction of this volume is optically thick.   

Some fraction of the light in the SED can still come from a blackbody or a disk, but any such contribution must be small so as to beat the limits above.  The most likely spectral regime for such an exception would be in the ultraviolet, where even a body with a small surface area can provide much luminosity.  Thus, a disk can still be hiding in the SED, with its light being swamped by nonthermal light, but the above limits mean that only a small fraction of the SED and of the accretion luminosity can come out as disk light.

At least three non thermal emission mechanisms can provide power law SEDs with roughly the correct slope.  The first is free-free emission.  The second is cyclotron emission by thermal electrons in a hot plasma, presumably near the white dwarf and fed by the accretion.  The third is from Compton scattering of low-energy seed photons by hot electrons in a surrounding plasma.

Nonthermal processes lack the simplicity of blackbody emission.  This means that any calculations (e.g., energy budgets, SED shape) depend on the details of the model, with these always being very complex, dependent on various unknown parameters, and largely unsolved.  As such, this paper cannot make any useful predictions for the non thermal emissions, and it is easy to imagine a variety of scenarios.

\subsection{The White Dwarf Magnetic Field}

Williams (1989), Oksanen \& Schaefer (2011), and Schaefer \& Collazzi (2010) have all pointed to the T Pyx system as having a highly magnetized white dwarf that channels the accretion flow onto a polar cap.  The first reason is that T Pyx shows photometric oscillations that vary in time with a period near to, but not equal to, that of the orbital period.  This is the hallmark of magnetic systems, and this one property has been used to identify many systems as magnetic.  For examples amongst novae, see RW UMi (Tamburini et al. 2007), V4633 Sgr (Lipkin \& Leibowitz 2008), V4745 Sgr (Dobrotka et al. 2006), V4743 Sgr (Kang et al. 2006), V697 Sco (Warner \& Woudt 2002), and V1495 Aql (Retter et al. 1998).  (But this is not a forcing argument, as there are other means to make non-orbital photometric periodicities, for example, superhumps.)  The second reason is that T Pyx is a quintessential example of the V1500 Cyg stars, wherein post-eruption nova are greatly brighter than pre-eruption novae while fading slowly for the next century or so.  All eight known cases have relatively short orbital periods, long-lasting supersoft emission, and highly magnetic white dwarfs; with all these uncommon shared properties likely forming a suite of requirements to get the bright post-eruption.  Schaefer \& Collazzi present a mechanism by which nuclear burning is sustained on the surface of the white dwarf (hence irradiating the companion star and driving the high accretion rate) by the magnetic field channelling the accretion onto a polar cap where the {\it local} accretion rate is high enough to sustain the nuclear burning.  (But this is not a strong reason, because there are other possible mechanisms that can produce the post-eruption irradiation-induced high mass transfer that characterizes the V1500 Cyg behavior.)  The third reason is that T Pyx has one of the highest fluxes in the He II $\lambda$4686 line, and this has been taken as pointing to magnetic accretion as the only way to get a high excitation luminosity (Williams 1989; Diaz \& Steiner 1994).  (But this is not a strong reason because little study or explanation of the connection, bright helium emission to high magnetic fields, has been made, much less proven to be unique.)  The fourth reason is that we have already proven that the emission from T Pyx cannot be from a disk of any type (see Section 5.2.4), and the only way to get the material to the surface of the white dwarf without a disk is through some sort of a magnetic accretion column.  (This is not a strong argument because perhaps the disk light is being swamped by nonthermal light from some other source in the system.)

The existence of a strong magnetic field would be helpful for explaining how the optical light curve of T Pyx can have a 0.15 mag amplitude modulation where the minimum covers about half an orbit, despite having an inclination of roughly 10$\degr$.  With T Pyx being seen essentially face-on, there can be no eclipses, while any illumination or irradiation effects must be small ($\cos[10\degr]=0.985$).  So there must be some asymmetry in the system that is tilted with respect to the orbital plane.  The obvious possibility, and we cannot think of any other possibility, is to invoke a tilted magnetic field in the white dwarf.  Nonthermal cyclotron emission can easily have a mild beaming pattern that changes by 15\% for a 20$\degr$ shift in viewing angle.  For example, maybe the accretion column is pointing far away from the normal to the orbital plane, or the magnetic field suffuses some cloud near the white dwarf with cyclotron emission preferential in some direction.  For this idea to work, the rotation period of the white dwarf must be exactly tied to the orbital period, because the photometric period is tied to the radial velocity period.  Again, any real calculations (e.g., beaming patterns and degree of polarization) are model dependent, complex, and vary with unknown system parameters.

The reality of the high magnetic field for T Pyx is not proven, and indeed there are troubles with the basic idea.  First, a plausible standard of proof is to measure significant circular polarization that changes with the orbital period, and T Pyx has no measured polarization.  (But this is not evidence, because no reports have been published yet with circular polarization measures around an orbit.  The linear polarization measures reported by Eggen et al., 1967, are for the time around peak brightness when the outer shell was shrouding the inner binary.)  Second, Patterson et al. (2012) have argued that the original non-orbital periodicity reported by Patterson et al. (1998) has never been seen to re-appear, so ``without confirmation, its evidentiary value must be reckoned as weak".  (This argument is weak because they are not denying their own data, and there is no requirement for the signal to be frequently present.)  Third, they further argue that the V1500 Cyg stars with orbital periods below the period gap (GQ Mus, V1974 Cyg, CP Pup, and RW UMi) have their non-orbital photometric periodicities of low phase stability, hence pointing to a superhump phenomenon instead of a magnetic white dwarf.  (This is a weak argument because if the earlier non-orbital signals had been from superhumping, then superhumps should have been repeatedly seen and clearly identified in their long photometric records.  This argument is also weak because these same stars have been identified as being highly magnetic systems on independent grounds; see Diaz \& Steiner 1994, Diaz \& Steiner 1991, Chochol 1999.)  Pointedly, Patterson et al. (2012) agree with Schaefer \& Collazzi that the V1500 Cyg stars with orbital periods {\it above} the period gap have high magnetic fields, so this discussion is just about the short period systems like T Pyx.  Fourth, the spectra of T Pyx show double peaked emission lines, with this shape characteristic of disks with a bright inner region.  (But this argument is weak because other structures, like pairs of accretion columns, can also produce the observed line profiles.)  Taking all the arguments pro-and-con together, we see no decisive case either way.  Until resolved, we judge that the existence of a high magnetic field on the T Pyx white dwarf is  reasonable yet undecided.

Fortunately, we already have excellent polarization data for T Pyx, taken in February 2013 with {\it FORS} on {\it UT1} at the {\it VLT} on Cerro Paranal in Chile.  This observing run has S. Katajainen as P.I. plus B. E. Schaefer and A. Oksanen as Co.I.s.  The answer to whether T Pyx is polarized, and hence whether its white dwarf is highly magnetized, is not yet known, but an answer will come soon.

\subsection{The 44 Year Inter-Eruption Interval}

Schaefer (2005) and Schaefer et al. (2010a) made a quantitative physics-based prediction that the date of the next T Pyx eruption would be in the year 2052 or much later.  The first key idea is that the accretion rate can be measured from the blue flux, so a B-band light curve between eruptions can be used to determine the average accretion rate.  This is certainly true for all emission mechanisms and structures.  For any particular system over a relatively restricted range of $\dot{M}$, we can represent or approximate the relation as a power law.  The general relation will be 
\begin{equation}
\dot{M} = \dot{M}_{16} F_B^E = \dot{M}_{16} 10^{-0.4(B-16)E},
\end{equation}
where $\dot{M}$ is the accretion rate, $\dot{M}_{16}$ is the accretion rate when T Pyx has a B magnitude of 16.0, $B$ is the B-band magnitude, $F_{B}$ is the B-band flux in units of the the flux when B=16, and $E$ an exponent representing the physics of the light sources in T Pyx.  The $E$ value will vary from source-to-source, depending on the structure of the accretion flow, the emission mechanism, and the wavelength of observation.  The use of the B=16 level is arbitrary (any other magnitude level could have been adopted), but this choice is convenient for T Pyx.  The second key idea is that the total mass accreted between eruptions can be estimated from the accretion rate averaged over the inter-eruption interval of $\Delta T$, with 
\begin{equation}
\Delta M_{acc} = \langle \dot{M} \rangle \Delta T.
\end{equation}
Here, the accretion rate values from equation 9 are averaged over all available blue magnitudes.  The third key idea is that the trigger mass will be close to the mass accreted between eruptions: 
\begin{equation}
M_{trigger} = \Delta M_{acc}.
\end{equation}
The fourth key idea is that a given recurrent nova system should have a unique trigger mass that will be the same for all eruptions.  The trigger mass can only be a function of the white dwarf mass, radius, magnetic field, and the composition of the accreted material, and these vary from RN to RN, but not from eruption to eruption.  This can be expressed as 
\begin{equation}
M_{trigger1} = M_{trigger2} = M_{trigger3} = \ldots,
\end{equation}
where the numerical subscript indicates different eruptions.  Putting these equations together, we see that $\langle F_B^E \rangle\Delta T$ should be a constant from eruption to eruption.  For intervals where the RN is bright on average, then the accretion rate will be high, the trigger mass will be accumulated relatively fast, and the inter-eruption interval will be relatively short.  For intervals where the RN is faint on average, the time between eruptions will be long.  For both T Pyx and U Sco (Schaefer 2005; Schaefer et al. 2010b), the blue flux was seen to vary with the short inter-eruption intervals having a bright quiescence (and thus a high accretion rate) and with the long inter-eruption intervals having a faint quiescence (and thus a low accretion rate).  We now can determine the current inter-eruption time interval as 
\begin{equation}
\Delta T_3 = \Delta T_2 \langle 10^{-0.4(B-16)E} \rangle _2 / \langle 10^{-0.4(B-16)E} \rangle _3.
\end{equation}  
The subscripts on the brackets indicate an average over the indicated inter-eruption interval.  

We now have a means of predicting the date of the next eruption by simply measuring the historical light curve along with some assumption for the dependence of the quiescent brightness on the accretion rate.  By calibrating with previous inter-eruption intervals, the various constants of proportionality cancel out, and we need never know the trigger mass, accretion rate, or the distance.  Schaefer (2005) predicted that the next eruption of U Sco would be in 2009.3$\pm$1.0, and this formed the basis for a large international collaboration of researchers at all wavelengths designed to discover and then intensively observe the eruption.  With the discovery of the eruption in 2010.1 by B. Harris and S. Dvorak, this advance preparation allowed the U Sco eruption to become the all-time best observed nova event. Schaefer (2005) also predicted that T Pyx would next erupt in 2052$\pm$3, while Schaefer et al. (2010a) extended this date to the year 2225 or much later. 

This prediction for T Pyx was refuted by the 2011 eruption.  So the task now is to understand {\it why} the prediction went so far wrong.  The basic logic of the prediction is simple and easy.  The light curve is confident and the mass for the trigger threshold really has to be constant.  About the only place that could be wrong is the conversion from the B-band magnitude to an accretion rate.  Schaefer (2005) calculated that the accretion rate is proportional to the square of the blue flux (i.e., $E=2.0$), with this exponent being derived from a model for a simple accretion disk with the parameters for T Pyx.  (For U Sco with its much larger disk, the exponent is 1.5.)  So this exponent is the likely cause of the wrong prediction.

The exponent $E$ is merely a description of the accretion-flux relation for a relatively small range of $\dot{M}$ for a given star and wavelength.  The $E$ value varies substantially.  We have already seen how $E=2.0$ and $E=1.5$ for the standard disk models of T Pyx and U Sco respectively (Schaefer 2005).  For T Pyx, we have repeated the calculations for a wide range of disk models, where a variable accretion rate goes through a disk with the inner and outer radii of the disk widely varying with and without irradiation on the disk, while we find that $E$ for the B-band varies from 1.0 to 3.1, with the higher values going for the higher accretion rates and the lower values going for disks with severe inner truncation.  The 1.21 M$_{\odot}$ models of Wade \& Hubeny (1998) have exponents of 1.31 and 1.41 for 1500\AA~ and 2000\AA~ at the highest range of accretion.  Retter \& Leibowitz (1998) give a simplistic theoretical model ($L_{bol} \propto \dot{M}$, $F \propto L_{bol}$, so $F \propto \dot{M}$) that gives $E=1$, but this has not included the critical variation of the bolometric correction as the accretion rate changes.  With this wide variation in models, it is clear that the Schaefer (2005) value of $E=2.0$ could well be wrong.  In addition, it is now clear that we can expect no confident guidance from models or theory, even within the framework of a disk model, and especially since the disk does not dominate the blue light.  Thus, we can only turn to empirical evidence regarding $E$ for T Pyx.

If the exponent is around 1.0, then the predicted date would have been close to 2011.  This is the case where the accretion rate is simply proportional to the B-band flux.  The quiescent magnitude from 1944 to 1967 was 14.72, while from 1967 to 2005 it was 15.49 (Schaefer 2005), for a magnitude difference of 0.77 mag, corresponding to a ratio of B-band flux of close to 2.0.  For the accretion rate being proportional to the B-band flux, the average accretion rate must then also have a ratio of close to a factor of 2.0, and the post-1967 interval must be 2.0 times that of the pre-1967 interval.  This would imply an eruption 2.0$\times$22 years after 1967, which predicts the next eruption in 2011.  So it seems that, somehow, T Pyx effectively has the B-band flux being proportional to the accretion rate (i.e., $E\approx 1$).

This fast and easy analysis can be substantially improved.  Using the quiescent light curve and the inter-eruption time intervals, the value of E that makes for the most constant $\langle F_B^E \rangle\Delta T$ from interval to interval can be found.  To determine the best fit exponent, we have taken our exhaustive light curves from each of the intervals (see Table 2), and formed averages of $F_B^E$ (see Table 3).  The only free parameter in this analysis is $E$.  The best estimate for $E$ will be that value which returns the most constant values for $\langle F_B^E \rangle\Delta T$ across all measured time intervals.  The values and the scatter will scale as $E$ changes, so we find the value of $E$ that has the least fractional scatter.  The smallest fractional scatter is near $E=1.13$, with 13\% RMS variations across the five intervals with measures.  With this, Table 3 has the value of $\langle F_B^{1.13} \rangle\Delta T$ in the last column, with a straight average of 71.  Table 3 also lists the values of $\langle F_B \rangle\Delta T$ (with a straight average of 64).

The scatter in $\langle F_B^E \rangle\Delta T$ is much larger than the formal uncertainties, and this indicates that our basic model is not exact.  So, at the 13\% level, the imperfect sampling of the light curve might be misrepresentative, the value of $E$ varies, or the trigger mass might vary from eruption to eruption.  For whatever reason, this means that the total uncertainty in $\langle F_B^E \rangle\Delta T$ is around 13\%, so for five independent measures the uncertainty on the average is around 6\%.  For predicting any single inter-eruption interval, the uncertainty will be 13\%.  Adopting a 13\% uncertainty, we can make a chi-square fit for $E$, with this giving a value of 1.13$\pm$0.14.  As such, we see that the five inter-eruption time intervals are consistent with $E=1$.

So we now have a plausible explanation for why T Pyx erupted in 2011.  The observed flux from T Pyx has been declining for the last 121 years, so the accretion rate has been falling and it takes longer times between eruptions to accumulate the nova trigger mass.  In the 22 year inter-eruption interval before the 1967 event, T Pyx was nearly a factor of two times brighter on average compared to the next inter-eruption interval, so the next inter-eruption interval should be roughly twice as long (i.e., around 44 years).  Therefore, the next eruption after 1967 should occur in 1967+44, which is to say in the year 2011.  A more accurate postdiction is to use the average value of $\langle F_B \rangle\Delta T = 64\pm9$ and $\langle F_B \rangle=1.47\pm0.011$ to get $\Delta T =43.6\pm 5.9$, which puts the eruption peak in 2010.7$\pm$5.9.  The uncertainty on this postdicted date is not small, so the close coincidence between the postdiction and the real eruption date is fairly surprising.

\subsection{The Nature of the Luminous Hot Source Irradiating the Companion Star}

Four different groups of researchers have concluded, on greatly different grounds, that T Pyx harbors a very hot and luminous source irradiating the companion star and surrounding shell.  Contini \& Prialnik (1997) require that the entire surface of the white dwarf emit with a temperature of 250,000$\degr$K (for a luminosity of over $10^{36}$ erg s$^{-1}$) so as to have the irradiation of the surrounding shell produce the observed spectrum.  Patterson et al. (1998) needed a supersoft source to produce the high observed luminosity as driven by the high observed accretion rate.  Knigge et al. (2000) invoked a wind-driven supersoft source so as to explain the central T Pyx mystery of how a binary with such a short orbital period (so that only General Relativistic effects should be driving a very very low mass transfer) can possibly have high enough of an accretion rate to make the system recurrent with a 12-44 year recurrence time scale.  Schaefer \& Collazzi (2010) needed the 1866 classical nova eruption to turn on the irradiation so as to make the high accretion rate that is secularly declining, both to explain the fading in quiescent brightness from B=13.8 in 1890 to B=15.7 in 2010 as well as to explain why the inter-eruption intervals have been getting longer and longer since 1890.  Thus, we see that the existence of a hot and luminous source explains many of the central mysteries of T Pyx, while no alternative idea has even been proposed.

Four mechanisms have been proposed to power the hot-luminous source near the surface of the white dwarf:  Contini \& Prialnik envisioned simply the surface of the white dwarf being very hot for unspecified reasons, likely due to residual heat from the eruptions.  Patterson et al. have their supersoft source powered by nuclear burning of the accreted material on the surface of the white dwarf.  Knigge et al. have the mass transfer rate sufficiently high that the accretion energy powers the source.  Schaefer \& Collazzi have the accretion being funneled onto the polar cap of a highly magnetic white dwarf  so that the local accretion rate is sufficiently high to make for steady hydrogen burning, and this nuclear power is what runs the hot-luminous source.  Importantly, uncertainties in the mechanism should not hide the broad agreement about what is going on in T Pyx (where the hot-luminous source is irradiating the companion, driving the high accretion rate, and determining the evolution of the system).  Also importantly, the mechanism operating on T Pyx need not be the same mechanism operating on either the other short-period V1500 Cyg star systems or the long-period V1500 Cyg star systems.

Despite the strong needs for a central hot-luminous source in T Pyx, the idea has the substantial problem that it has never been directly seen, and the X-ray limits put strong constraints on possible sources.  Selvelli et al. (2010) demonstrate that a blackbody with a temperature of 240,000$\degr$K ($kT=0.020$ keV) at 3 kpc with $10^{37}$ erg s$^{-1}$ luminosity is nearly two orders-of-magnitude above the threshold for the {\it XMM} limit.  This explicitly excludes the default cases proposed by Contini \& Prialnik, Patterson et al., and Knigge et al.  (Schaefer \& Collazzi did not propose any specific parameters for their hot-luminous source.)  So for the hot-luminous source idea to pass, there must be some device to hide the X-ray flux.  Here are four possibilities:  First, the very hot nuclear burning region can be covered with a shroud of gas that is substantially colder, so that the luminosity comes out in the unseen far ultraviolet, with essentially no flux above 0.2 keV.  Second, the feedback between the accretion luminosity and the accretion rate is necessarily unstable, and the accretion rate is seen to be falling, so the {\it current} luminosity might well be below the {\it XMM} threshold.  Third, perhaps most of the heating of the companion star atmosphere occurred soon after the 1866 classical nova eruption, and the high accretion rate has been driven by this latent heat, so the hot-luminous source has long been turned off.  (Determining the time scale for restoring the companion star to its pre-1866 atmosphere is a complex unsolved problem which depends greatly on the assumed model parameters.)  Fourth and simplest, maybe the hot-luminous source is simply cooler than an effective temperature of $\sim$50,000$\degr$K (e.g., the 34,000$\degr$K give by Gilmozzi \& Selvelli 2007), and all would be well.  Given these many reasonable resolutions, we judge that the X-ray limit is not a serious problem for the hot-luminous source idea.

Nevertheless, the particular mechanisms have substantial problems.

The possibility that the hot-luminous source is powered by steady nuclear burning of accreted material (Patterson et al. 1998; Schaefer \& Collazzi 2010) has the problem that the burning itself gets rid of the very fuel needed for the recurrent nova eruptions.  That is, with the hydrogen being burnt up by the supersoft source, there will be no hydrogen left to power the later nova events.  (Importantly, this problem does {\it not} apply to the other V1500 Cyg stars, because they have no nova event until long after the supersoft source has turned off and normal hydrogen-rich material has had plenty of time to accumulate on the surface of the white dwarf.  So this is a problem for T Pyx alone.)  A possible dodge to this conclusion is that the hydrogen might only partially burn, with enough hydrogen accumulating, perhaps away from a magnetic pole cap, to allow for a normal eruption.

The possibility that the hot-luminous source is powered by the accretion energy (Knigge et al. 2000) has the problem that the current accretion rate is very close to the limit for steady hydrogen burning, so that the rate over a century ago would certainly be above the threshold.  The accretion rate is $1.7-3.5\times 10^{-7}$ M$_{\odot}$ yr$^{-1}$, as measured from the steady period change ($\dot{P}$) in the O-C diagram for the time interval 1968 to 2011.  But back before the 1890 eruption, the $\langle F \rangle$ value was 5.7$\times$ brighter, so with $E=1$ the accretion rate must have been 5.7$\times$ larger with $\dot{M}=2.6-5.5 \times 10^{-6}$ M$_{\odot}$ yr$^{-1}$.  For a 1.3 M$_{\odot}$ white dwarf, the threshold for steady hydrogen burning is $2-3 \times 10^{-7}$ M$_{\odot}$ yr$^{-1}$ (Nomoto 1982; Yaron et al. 2005; Nomoto et al. 2007; Shen \& Bildsten 2007).  Thus, before the 1966 eruption, T Pyx was certainly accreting at a rate that places it in the steady hydrogen burning regime.  But T Pyx was undergoing recurrent nova eruptions from 1890 to 1966, so there is something wrong here.  A possible solution to this dilemma is that the highly variable and fast changing accretion history makes the thermal structure of the white dwarf far from equilibrium and colder than expected for its current accretion rate, and this will effectively shift the threshold for steady hydrogen burning to some larger limit.  That is, for a million years before 1866, the white dwarf's thermal structure equilibrated to a very low $\dot{M}$ state with no eruptions, and then this equilibrium will be greatly perturbed by the various eruptions and the very high accretion rate, thus resulting in an out-of-equilibrium white dwarf that is much colder than expected for its current state as a RN.  It will take detailed physics calculations to test the size of the effect of the variable accretion rate and eruption history. 

In all, we judge that the case for the existence of the hot-luminous central source to be persuasive but not proven, while the details of the mechanism that power it are not known, even while reasonable possibilities are known.

\section{A Full Model For the T Pyx System}

T Pyx has a fairly complicated history that has been pieced together from a wide variety of data.  It might be useful to have one consistent picture explicitly collected in one place.  Here, we present a largely chronological summary based primarily on this paper, Schaefer (2010), Schaefer \& Collazzi (2010), and Schaefer et al. (2010a).

For $\sim$750,000 years before 1866, T Pyx was a nondescript short-period cataclysmic variable (possibly with a highly magnetic white dwarf) with an accretion rate of around $4 \times 10^{-11}$ M$_{\odot}$ yr$^{-1}$ and $B=18.5$ mag.  After this long and slow accumulation of mass on the white dwarf, a normal classical nova eruption occurred in the year 1866$\pm$5, reaching perhaps $B=5$ mag.  This eruption sent out a smooth shell with a mass of $\sim10^{-4.5}$ M$_{\odot}$ (larger than the accumulated mass on the white dwarf) at a velocity of 500-715 km s$^{-1}$ (for the traditional distance of 3500 pc).  This eruption irradiated the nearby companion star (which is very close due to the short orbital period) so as to drive a high rate of mass being pushed through the L1 point in the Roche lobe.  With some combination of the high accretion energy and steady hydrogen burning (perhaps on a magnetic polar cap), the white dwarf remained as a hot high-luminosity source to continue the irradiation and keep the high $\dot{M}$ at a nearly self-sustaining rate.  After the 1866 eruption was long over, T Pyx stayed as a bright high-accretion system, much in contrast to its state a few years previously, which is to say that it is a V1500 Cyg star.  With its newfound high accretion rate, the white dwarf accumulated material quickly, and with it being near the Chandrasekhar mass, the system started producing recurrent novae events.  Each successive recurrent nova eruption sent out a fast (2000 km/s) shell of relatively low mass, with these shells ramming into the slower 1866 shell causing thousands of knots to form by Rayleigh-Taylor instabilities.  The feedback loop of irradiation and accretion is not stable, so the accretion rate must be declining, with the typical time scale for such cases being a century or so.  With this, the accretion rate must decline, the brightness of the system must decline (as observed to fade from B=13.8 in 1890 to B=15.7 in 2011), and the overall recurrent nova phase (starting in 1866) must have a brief duration.  Another consequence of the falling accretion rate is that the duration of the inter-eruption intervals must increase.

Two or three requirements are needed for this history to apply to T Pyx: a white dwarf near the Chandrasekhar mass, a very short orbital period, and possibly a significant magnetic field on the white dwarf.  The combination of the first two requirements is rare, so there will be few such stars in our Milky Way and the durations of their recurrent nova phase will be short, perhaps one-to-a-few centuries out of millions of years.  Nevertheless, their frequent and bright eruptions will call attention to these systems from far across the galaxy.

Although T Pyx is a system with material falling at a very high rate onto a white dwarf near the Chandrasekhar mass, it is not a progenitor of a Type Ia supernova.  First, the mass gain/loss in the system is dominated by the classical nova events, and these always eject more mass than has been accreted.  Second, the recurrent nova events also eject more mass than they accrete, as we have just learned from the measured period changes across the 2011 eruption of T Pyx, the 2010 eruption of U Sco, and the 2000 eruption of CI Aql.  Third, systems like T Pyx are too rare and their lifetimes too long to provide enough Type Ia supernovae.  (This statement is not true for the long-period RNe.)  Fourth, half of all RNe are neon novae, with this making Type Ia supernovae impossible.  So, not only is T Pyx not a progenitor, but none of the RNe are progenitors.

The irradiation of the low mass companion star causes a high accretion rate, which is measured to be $1.7-3.5\times10^{-7}$ M$_{\odot}$ yr$^{-1}$ as based on the the dynamics of the mass transfer.  This rate is close to the threshold for which a white dwarf will steadily burn the accreted hydrogen.  The mechanism to power the hot-luminous source is not known, but published possibilities include using the gravitational energy of the accreted material or the nuclear energy from steady hydrogen burning, while the material may or may not be funneled onto a magnetic pole of the white dwarf.  The hot-luminous source cannot have an effective temperature over $\sim$50,000$\degr$K so as to avoid the limits from the X-ray flux.  The ultraviolet-optical-infrared spectral energy distribution follows closely to a power law with $f_\nu \propto \nu^{1.0}$, although the relatively narrow ultraviolet region is tilted to more like $f_\nu \propto \nu^{1/3}$.  The bulk of the flux cannot come from an accretion disk, nor any superposition of blackbody spectra, so most of the accretion light is coming out from some nonthermal emission process (free-free, cyclotron, or Compton scattering) out of a volume within the Roche lobe of the white dwarf that is largely optically thin.  With the realization that the optical light is not from a disk, we can break out of the idea that lead to the incorrect prediction that the next eruption was far into the future.  Instead, we have something close to $F_B \propto \dot{M}$.  As such, the 22 year inter-eruption interval before 1967 (with $\langle B \rangle = 14.70$ mag) has a flux roughly a factor of two brighter than the inter-eruption interval after 1967 (with $\langle B \rangle = 15.59$ mag), so its duration should be roughly 2$\times$22 years and the next eruption around the year 1967+44=2011.

Based on the prior average inter-eruption interval, many people expected an imminent nova event starting in the mid-1980's.  No eruption could have been missed (say, due to solar gaps) because the T Pyx events have a long duration (roughly 230 days, with $t_3=62$ days) and a bright peak (B=6.7 mag) such that the nightly observations throughout the entire observing season would certainly have caught any eruption.  Nevertheless, many observers in the professional and amateur communities persevered by taking many time series and nightly monitoring for decades on end.  From 1986-2011, the orbital period had a steady increase at $P/\dot{P}=313,000$ years, although a small deviation appears from 2008-2011 as expected for the small observed drop in accretion luminosity over this time.  In the several years before 2011, T Pyx averaged V=15.5 mag, and was never seen brighter than V=15.0.

It came as a surprise to Stubbings when T Pyx appeared at V=14.5 on UT 2011 April 5.51, and he realized that this might presage an eruption.  Instead, he saw a unique rise and fall in brightness with a peak at V=14.4 over several days.  This pre-eruption rise started just 18 days before the eruption.  Only four anticipatory rise have ever been seen (T Pyx, V533 Her, V1500 Cyg, and T CrB), with all being greatly different from each other, and all with zero theoretical understanding.  

As part of his long series of monitoring of recurrent novae, Linnolt saw T Pyx at V=14.5 on UT 2011 April 13.36 and realized that this was unusually bright.  On the next night at UT 2011 April 14.29, Linnolt saw T Pyx at V=13.0 and immediately recognized the eruption.  Within 15 minutes the world had been alerted, Plummer and Kerr soon confirmed the eruption, and Nelson, Carstens, and Streamer started time series in BVI.  For the first two days, the magnitude during this initial rise is closely represented as that of a uniformly expanding shell.  The steady and frequent monitoring, principally by amateur observers, fast recognition and notification, and speedy response around the world all make for T Pyx being the first and only nova eruption where the light curve was followed throughout the entire initial rise.  Indeed, while all prior novae events have returned only a few single magnitudes in the late part of the fast initial rise, for T Pyx, we now have a complete record of magnitudes and colors over the entire initial rise with high time resolution.

~

Throughout this work, the AAVSO provided charts, comparison sequences, data archiving, and alert notices.  This work is supported under grants from the National Science Foundation and from NASA as part of the {\it GALEX} program.  The NASA Galaxy Evolution Explorer, {\it GALEX}, is operated for NASA by the California Institute of Technology under NASA contract NAS5-98034.  AUL acknowledges support from NSF grants MPS 75-01890 and AST 0803158, and from AFOSR grants 77-3218 and 82-0192.  AP is supported by the Edward Corbould Research Fund and the Astronomical Association of Queensland.  MS wishes to thank the AAVSO for the loan of the SBIG 402 camera.  WGD thanks Arne Rosner at Global Rent-A-Scope in Moorook, South Australia.

{}

\begin{deluxetable}{lllllll}
\tabletypesize{\scriptsize}
\tablewidth{0pc}
\tablecaption{Pre-eruption Rises}
\tablehead{\colhead{Nova} & \colhead{Nova Class\tablenotemark{a}\tablenotemark{b}} & \colhead{$V_{peak}$\tablenotemark{a}} & \colhead{$V_Q$\tablenotemark{a}} & \colhead{$\Delta V_{rise}$\tablenotemark{c}} & \colhead{$T-T_{start}$ (days)\tablenotemark{c}\tablenotemark{d}} & \colhead{Rise Morphology\tablenotemark{c}}}
\startdata
T Pyx	 & 	P(62), RN, short-P$_{orb}$, shell	 & 	6.4	 & 	15.5	 & 	1.1	 & 	 -18 to -4	 & 	`Bump'	 \\
V533 Her	 & 	S(43), CN, IP, shell	 & 	3.0	 & 	15.0	 & 	1.7	 & 	 -550 to 0	 & 	Exponential rise	 \\
V1500 Cyg	 & 	S(4), neon CN, Polar, shell	 & 	1.9	 & 	20.5\tablenotemark{c}	 & 	7.0	 & 	$<$-23 to 0	 & 	Fast and bright rise	 \\
T CrB	 & 	S(6), RN, Red Giant donor	 & 	2.5	 & 	10.5	 & 	1.4	 & 	 -3200 to -220	 & 	`Bump'	 \\\enddata
\tablenotetext{a}{Data from Strope, Schaefer, \& Henden (2010).}
\tablenotetext{b}{Class P has a plateau in the light curve.  Light curve class S has a smooth monotonic decline.  The parentheses contains the $t_3$ decline rate.  CN indicates a classical nova with no known recurrences.  IP indicates an intermediate polar system with a highly magnetized white dwarf.}
\tablenotetext{c}{Data from Collazzi et al. (2009) and this paper.}
\tablenotetext{d}{$T-T_{start}$ is the time from the beginning and end of the pre-eruption rise to the start of the eruption.}\label{Table1}
\end{deluxetable}

\begin{deluxetable}{llll}
\tabletypesize{\scriptsize}
\tablewidth{0pc}
\tablecaption{T Pyx in quiescence (1890-2011) and the initial rise of the 2011 eruption}
\tablehead{\colhead{Julian Date} & \colhead{Band} & \colhead{Magnitude} & \colhead{Source}}
\startdata
2411491	 & 	B	 & 	13.80	 $\pm$ 	0.15	 & 	HCO plates	\\
2411493.6	 & 	B	 & 	13.80	 $\pm$ 	0.30	 & 	HCO plates	\\
2412211	 & 	B	 & 	14.10	 $\pm$ 	0.15	 & 	HCO plates	\\
2412236	 & 	B	 & 	14.30	 $\pm$ 	0.15	 & 	HCO plates	\\
2412936	 & 	B	 & 	14.30	 $\pm$ 	0.15	 & 	HCO plates	\\
2413570	 & 	B	 & 	14.30	 $\pm$ 	0.15	 & 	HCO plates	\\
2413891	 & 	B	 & 	14.30	 $\pm$ 	0.15	 & 	HCO plates	\\
2414424	 & 	B	 & 	14.50	 $\pm$ 	0.15	 & 	HCO plates	\\
2414454	 & 	B	 & 	14.70	 $\pm$ 	0.15	 & 	HCO plates	\\
2414766	 & 	B	 & 	14.70	 $\pm$ 	0.15	 & 	HCO plates	\\
$\ldots$	 & 		 & 	 & 		\\
2455635.1653	 & 	Vis.	 & 	15.40	 $\pm$ 	0.15	 & 	Stubbings	\\			
2455641.7291	 & 	ASAS-V	 & 	15.16	 $\pm$ 	0.09	 & 	ASAS	\\			
2455643.7340	 & 	ASAS-V	 & 	15.59	 $\pm$ 	0.10	 & 	ASAS	\\			
2455645.7339	 & 	ASAS-V	 & 	15.03	 $\pm$ 	0.09	 & 	ASAS	\\			
2455647.7350	 & 	ASAS-V	 & 	15.12	 $\pm$ 	0.08	 & 	ASAS	\\			
2455649.0903	 & 	Vis.	 & 	15.30	 $\pm$ 	0.15	 & 	Stubbings	\\			
2455650.0708	 & 	Vis.	 & 	15.40	 $\pm$ 	0.15	 & 	Stubbings	\\			
2455650.8194	 & 	Vis.	 & 	15.20	 $\pm$ 	0.15	 & 	Linnolt	\\			
2455652.1556	 & 	Vis.	 & 	15.40	 $\pm$ 	0.15	 & 	Stubbings	\\			
2455652.5797	 & 	ASAS-V	 & 	15.18	 $\pm$ 	0.10	 & 	ASAS	\\			
2455654.5972	 & 	ASAS-V	 & 	15.00	 $\pm$ 	0.09	 & 	ASAS	\\			
2455656.6210	 & 	ASAS-V	 & 	14.69	 $\pm$ 	0.08	 & 	ASAS	\\			
2455657.0104	 & 	Vis.	 & 	14.50	 $\pm$ 	0.15	 & 	Stubbings	\\			
2455657.0153	 & 	Vis.	 & 	14.50	 $\pm$ 	0.15	 & 	Stubbings	\\			
2455657.0222	 & 	Vis.	 & 	14.50	 $\pm$ 	0.15	 & 	Stubbings	\\			
2455657.0285	 & 	Vis.	 & 	14.40	 $\pm$ 	0.15	 & 	Stubbings	\\			
2455657.0361	 & 	Vis.	 & 	14.50	 $\pm$ 	0.15	 & 	Stubbings	\\			
2455657.0424	 & 	Vis.	 & 	14.60	 $\pm$ 	0.15	 & 	Stubbings	\\			
2455657.0993	 & 	Vis.	 & 	14.70	 $\pm$ 	0.15	 & 	Stubbings	\\			
2455657.1354	 & 	Vis.	 & 	14.70	 $\pm$ 	0.15	 & 	Stubbings	\\			
2455657.9667	 & 	Vis.	 & 	14.70	 $\pm$ 	0.15	 & 	Stubbings	\\			
2455658.6256	 & 	ASAS-V	 & 	15.07	 $\pm$ 	0.08	 & 	ASAS	\\			
2455659.0007	 & 	Vis.	 & 	14.80	 $\pm$ 	0.15	 & 	Stubbings	\\			
2455660.6229	 & 	ASAS-V	 & 	15.26	 $\pm$ 	0.08	 & 	ASAS	\\			
2455662.0417	 & 	Vis.	 & 	15.00	 $\pm$ 	0.15	 & 	Stubbings	\\			
2455664.8590	 & 	Vis.	 & 	14.50	 $\pm$ 	0.15	 & 	Linnolt	\\			
2455665.4752	&	B	&	12.51	 $\pm$ 	0.01	 & 	Schaefer	\\			
2455665.4986	&	B	&	12.50	 $\pm$ 	0.01	 & 	Schaefer	\\			
2455665.7931	 & 	Vis.	 & 	13.00	 $\pm$ 	0.15	 & 	Linnolt	\\
2455665.8847	 & 	Vis.	 & 	12.20	 $\pm$ 	0.15	 & 	Plummer	\\			
2455665.9313	 & 	Vis.	 & 	12.00	 $\pm$ 	0.15	 & 	Plummer	\\			
2455665.9410	 & 	Vis.	 & 	11.30	 $\pm$ 	0.15	 & 	Kerr	\\			
2455665.9667	 & 	Vis.	 & 	11.50	 $\pm$ 	0.15	 & 	Plummer	\\			
2455665.9681	 & 	Vis.	 & 	11.10	 $\pm$ 	0.15	 & 	Kerr	\\			
2455665.9938	 & 	Vis.	 & 	11.10	 $\pm$ 	0.15	 & 	Kerr	\\			
2455665.9944	 & 	Vis.	 & 	11.40	 $\pm$ 	0.15	 & 	Stubbings	\\			
2455665.9973	 & 	V	 & 	11.54	 $\pm$ 	0.01	 & 	Nelson	\\			
2455665.9978	 & 	V	 & 	11.55	 $\pm$ 	0.01	 & 	Nelson	\\			
2455665.9985	 & 	V	 & 	11.56	 $\pm$ 	0.01	 & 	Nelson	\\			
2455665.9990	 & 	V	 & 	11.56	 $\pm$ 	0.01	 & 	Nelson	\\			
2455665.9997	 & 	V	 & 	11.54	 $\pm$ 	0.01	 & 	Nelson	\\			
$\ldots$	 & 		 & 	 & 		\\
\enddata
\label{Table2}
{(This table is available in its entirety in a\\ machine-readable form in the
online journal.\\ A portion is shown here for guidance regarding\\ its form and
content.)}
\end{deluxetable}

\begin{deluxetable}{lllllll}
\tabletypesize{\scriptsize}
\tablewidth{0pc}
\tablecaption{T Pyx Inter-eruption Brightness and Flux}
\tablehead{\colhead{Interval} & \colhead{$\Delta T$ (yr)} & \colhead{Nights} & \colhead{$\langle B \rangle$ (mag)} & \colhead{$\langle F_B \rangle$} & \colhead{$\langle F_B \rangle  \Delta T$} & \colhead{$\langle F_B^{1.13} \rangle \Delta T$}}
\startdata
Pre-1890	&	\ldots	&	2	&	13.80	 $\pm$ 	0.13	&	7.59	 $\pm$ 	0.99	&	\ldots			&	\ldots			\\
1890-1902	&	11.86	&	12	&	14.38	 $\pm$ 	0.05	&	4.52	 $\pm$ 	0.20	&	54	 $\pm$ 	2	&	65	 $\pm$ 	3	\\
1902-1920	&	17.92	&	7	&	14.74	 $\pm$ 	0.05	&	3.20	 $\pm$ 	0.15	&	57	 $\pm$ 	3	&	67	 $\pm$ 	4	\\
1920-1944	&	24.62	&	68	&	14.88	 $\pm$ 	0.03	&	2.88	 $\pm$ 	0.08	&	71	 $\pm$ 	2	&	82	 $\pm$ 	2	\\
1944-1967	&	22.14	&	28	&	14.70	 $\pm$ 	0.04	&	3.35	 $\pm$ 	0.11	&	74	 $\pm$ 	2	&	87	 $\pm$ 	3	\\
1967-2011	&	44.33	&	336	&	15.59	 $\pm$ 	0.01	&	1.47	 $\pm$ 	0.011	&	65	 $\pm$ 	1	&	69	 $\pm$ 	1	\\
\enddata
\label{Table3}
\end{deluxetable}

\begin{deluxetable}{lllllll}
\tabletypesize{\scriptsize}
\tablewidth{0pc}
\tablecaption{T Pyx Minimum Times}
\tablehead{\colhead{Year} & \colhead{Observer} & \colhead{$\Delta$ (days)} & \colhead{HJD minimum} & \colhead{N} & \colhead{Linear O-C (days)} & \colhead{Quadratic O-C (days)}}
\startdata
1986.022	 & 	Warner\tablenotemark{a}	 & 	0.226	 & 	2446439.5121	 $\pm$ 	0.0035	 & 	-68378	 & 	0.1267	 & 	0.0073	 \\
1987.943	 & 	Szkody\tablenotemark{a}	 & 	0.091	 & 	2447141.0770	 $\pm$ 	0.0056	 & 	-59174	 & 	0.0960	 & 	0.0065	 \\
1988.347	 & 	Schaefer\tablenotemark{a}	 & 	1.038	 & 	2447288.5651	 $\pm$ 	0.0017	 & 	-57239	 & 	0.0844	 & 	0.0006	 \\
1988.840	 & 	Schaefer\tablenotemark{a}	 & 	0.154	 & 	2447468.9843	 $\pm$ 	0.0043	 & 	-54872	 & 	0.0737	 & 	-0.0033	 \\
1989.053	 & 	Bond\tablenotemark{a}	 & 	0.183	 & 	2447546.6591	 $\pm$ 	0.0039	 & 	-53853	 & 	0.0729	 & 	-0.0012	 \\
1989.537	 & 	Schaefer\tablenotemark{a}	 & 	0.196	 & 	2447723.4983	 $\pm$ 	0.0038	 & 	-51533	 & 	0.0649	 & 	-0.0030	 \\
1990.342	 & 	Buckley\tablenotemark{a}	 & 	0.459	 & 	2448017.3506	 $\pm$ 	0.0025	 & 	-47678	 & 	0.0611	 & 	0.0030	 \\
1990.348	 & 	Walker\tablenotemark{a}	 & 	0.193	 & 	2448019.6358	 $\pm$ 	0.0038	 & 	-47648	 & 	0.0595	 & 	0.0014	 \\
1996.112	 & 	Kemp\tablenotemark{b}	 & 	3.667	 & 	2450124.8300	 $\pm$ 	0.0005	 & 	-20030	 & 	0.0095	 & 	-0.0010	 \\
1996.220	 & 	Kemp\tablenotemark{b}	 & 	0.892	 & 	2450164.3930	 $\pm$ 	0.0009	 & 	-19511	 & 	0.0106	 & 	0.0006	 \\
1996.289	 & 	Kemp\tablenotemark{b}	 & 	0.328	 & 	2450189.5480	 $\pm$ 	0.0015	 & 	-19181	 & 	0.0106	 & 	0.0009	 \\
1996.352	 & 	Kemp\tablenotemark{b}	 & 	0.386	 & 	2450212.4910	 $\pm$ 	0.0014	 & 	-18880	 & 	0.0092	 & 	-0.0002	 \\
1997.031	 & 	Kemp\tablenotemark{b}	 & 	1.784	 & 	2450460.6090	 $\pm$ 	0.0007	 & 	-15625	 & 	0.0075	 & 	0.0010	 \\
1997.272	 & 	Kemp\tablenotemark{b}	 & 	0.686	 & 	2450548.4970	 $\pm$ 	0.0011	 & 	-14472	 & 	0.0055	 & 	-0.0001	 \\
2008.000	 & 	Richards\tablenotemark{c}	 & 	0.160	 & 	2454467.1391	 $\pm$ 	0.0015	 & 	36935	 & 	0.0334	 & 	-0.0016	 \\
2008.019	 & 	Richards\tablenotemark{c}	 & 	0.220	 & 	2454474.0788	 $\pm$ 	0.0013	 & 	37026	 & 	0.0364	 & 	0.0012	 \\
2008.058	 & 	Richards\tablenotemark{c}	 & 	0.292	 & 	2454488.1785	 $\pm$ 	0.0011	 & 	37211	 & 	0.0341	 & 	-0.0015	 \\
2011.099	 & 	Myers\tablenotemark{c}	 & 	0.148	 & 	2455598.7619	 $\pm$ 	0.0016	 & 	51780	 & 	0.0627	 & 	-0.0059	 \\
2011.118	 & 	Nelson\tablenotemark{c}	 & 	0.297	 & 	2455606.0039	 $\pm$ 	0.0011	 & 	51875	 & 	0.0631	 & 	-0.0057	 \\
\enddata
\tablenotetext{a}{Schaefer et al. (1992)}
\tablenotetext{b}{Patterson et al. (1998)}
\tablenotetext{c}{Table 2}
\label{Table4}
\end{deluxetable}

\clearpage
\begin{figure}
\epsscale{0.8}
\plotone{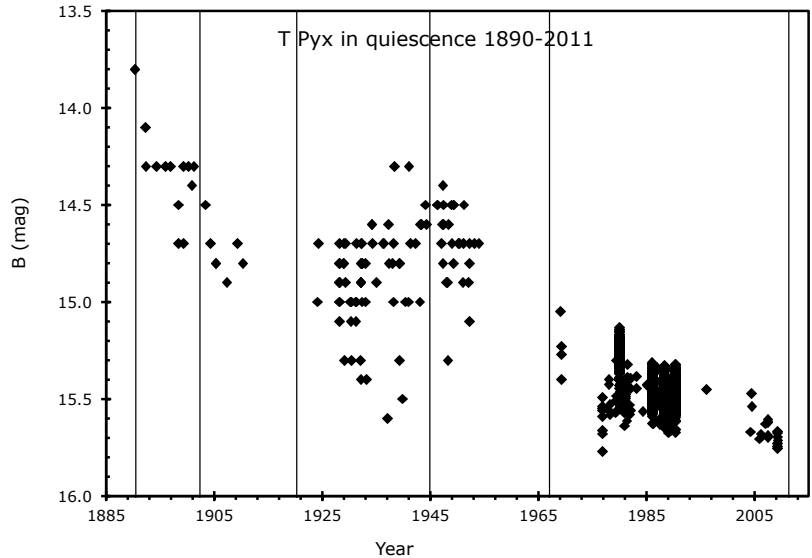}
\caption{
T Pyx in quiescence since 1890.  This light curve has all the B-band magnitudes from Table 2, with the vertical lines indicating the times of the six historical eruptions.  The key point is that T Pyx is suffering from a large and highly significant secular fade from B=13.8 in the quiescence before the 1890 eruption to B=15.7 just before the 2011 eruption.  This fairly steady decline is by 1.9 magnitudes, a factor of 5.7$\times$ in flux, which shows that the accretion rate has declined by at least a factor of 5.7 over this 122 year interval.  The cause for this decline is interpreted as due to a fading of the white dwarf after its 1866 classical nova eruption, with the white dwarf light irradiating and puffing up the companion star leading to weirdly high accretion rate in the system.  T Pyx appears to be going into hibernation now, although it is unclear how far down the accretion rate will ultimately fall to.  This falling accretion rate explains why the inter-eruption intervals have been increasing over time, simply because it takes longer and longer for the white dwarf to accumulate the necessary trigger mass.  This secular decline has the implication that T Pyx will not become a Type Ia supernova.}
\end{figure}

\clearpage
\begin{figure}
\epsscale{0.8}
\plotone{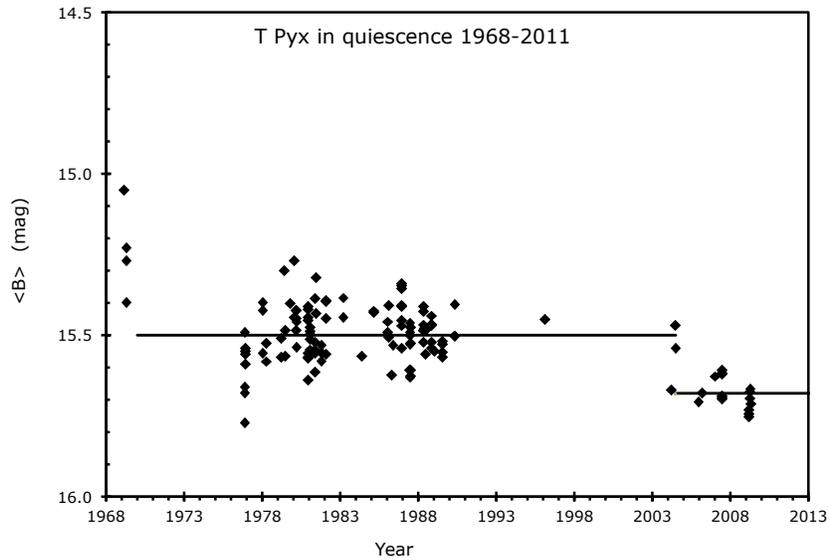}
\caption{
T Pyx in quiescence from 1968 to 2011.  The earliest points in this light curve, from 1969, show T Pyx in the late tail of its 1967 eruption.  The nights with time series photometry have been displayed with only the nightly average.  The B-band photometry from 1976 until 2004.5 shows a flat line, with the usual flickering, at $B=15.50\pm0.01$.  Both the B-band and V-band photometry show that T Pyx had an apparently sudden drop in brightness around the middle of 2004, fading to $B=15.68\pm0.01$.  The light from T Pyx is essentially all from the accretion process, so a sudden drop in brightness can only come from a sudden change in the accretion rate.  The steady orbital period change is proportional to the accretion rate, so this light curve implies that $\dot{P}$ will have a small step change starting in 2004.5.}
\end{figure}

\clearpage
\begin{figure}
\epsscale{0.8}
\plotone{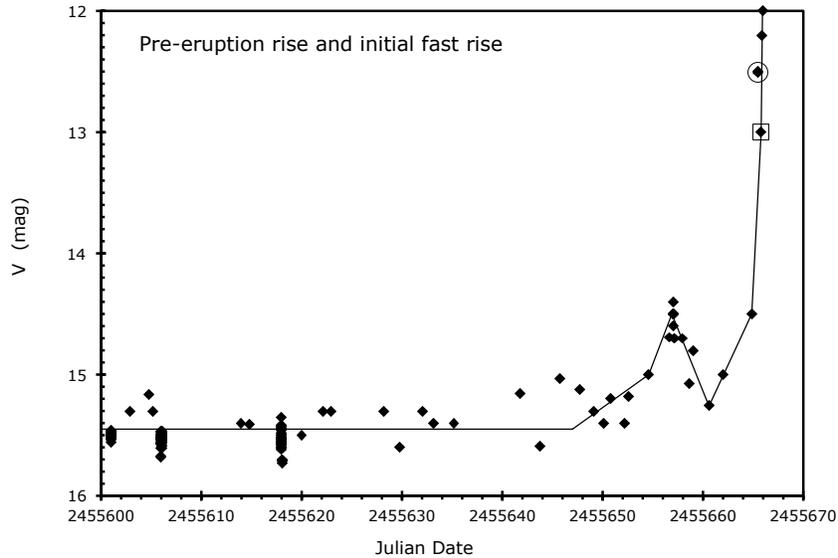}
\caption{
Pre-eruption rise.  T Pyx never appeared brighter than V=15.0 for thousands of days before the eruption, and then it suddenly brightened to V=14.4 mag just 9 days before the eruption.  The close time coincidence and the uniqueness of the bump demonstrate some causal connection between the bump and the eruption.  The decline before the very fast rise demonstrates that this significant bump in the light curve is not some early part of the thermonuclear runaway.  There is no understanding of the cause of the bump or why it anticipates the eruption.  All of the magnitudes on this plot are V-band magnitudes, except for the pair of B-band magnitudes circled in the upper right corner of the plot.  These two magnitudes are serendipitously taken 7.6 hours before the discovery of the eruption by Linnolt (indicated by the large square), with the earlier observations being 0.5 mag brighter than the later discovery.  We are left with either that the initial fast rise had a large amplitude dimming just hours before discovery or that T Pyx was extremely blue ($B-V= -1.0$) in its earliest moments.  Explaining the pre-eruption phenomena is now a premier challenge for nova theory.}
\end{figure}

\clearpage
\begin{figure}
\epsscale{0.8}
\plotone{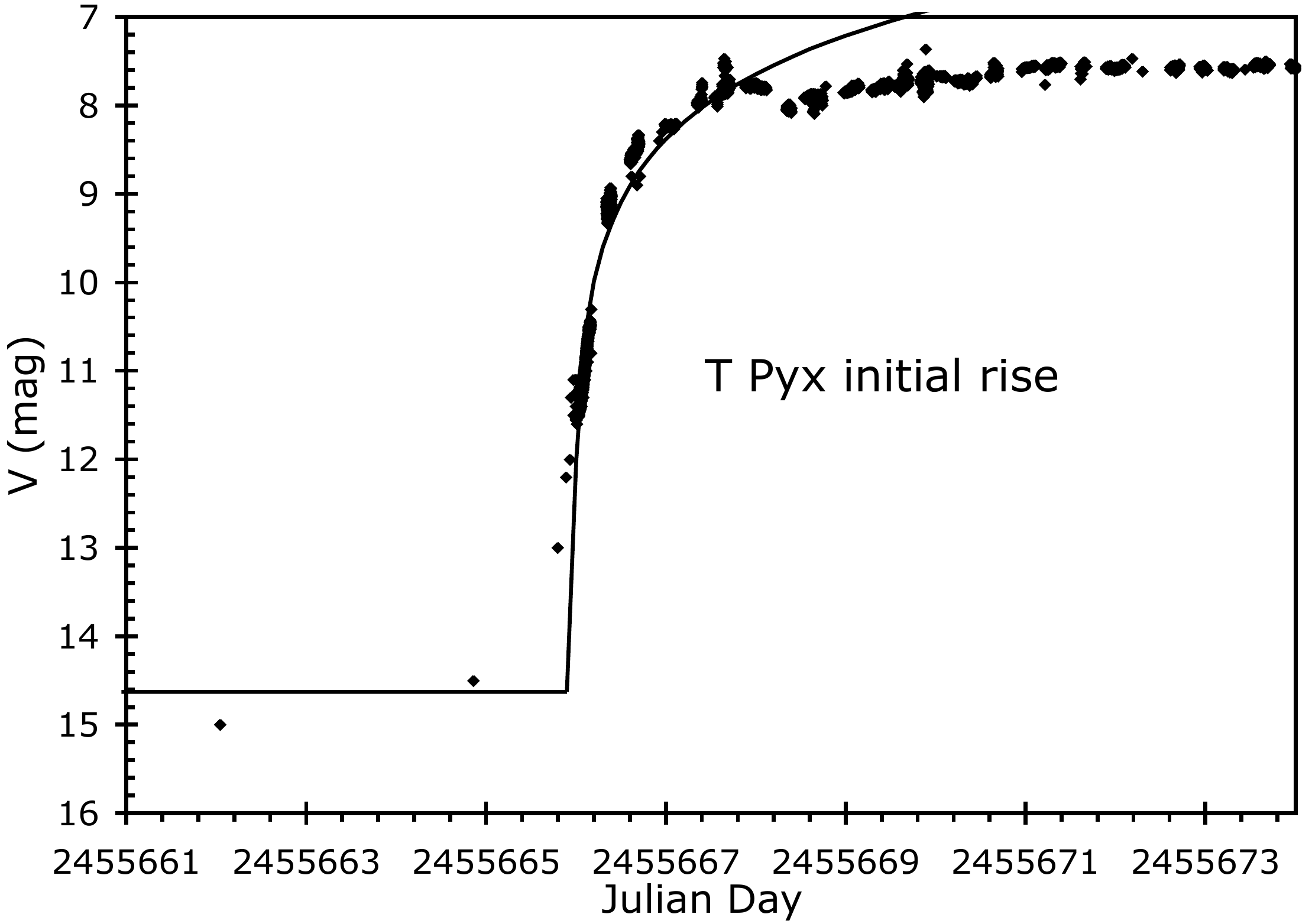}
\caption{
Initial rise.  T Pyx now has the only nova light curve where the brightness was followed over the entire initial rise.  For the first time, we see that the initial rise starts suddenly and is very fast.  The rise is as fast as 9 magnitudes per day.  A uniformly expanding shell is expected to rise fast in brightness followed by a substantial slowing (in magnitude units) after the first day, with the best fit model displayed by the curve in the figure.  We see a sudden small fading and break from the model about two days after the start, with this perhaps indicating the recession of the photosphere.}
\end{figure}

\clearpage
\begin{figure}
\epsscale{0.8}
\plotone{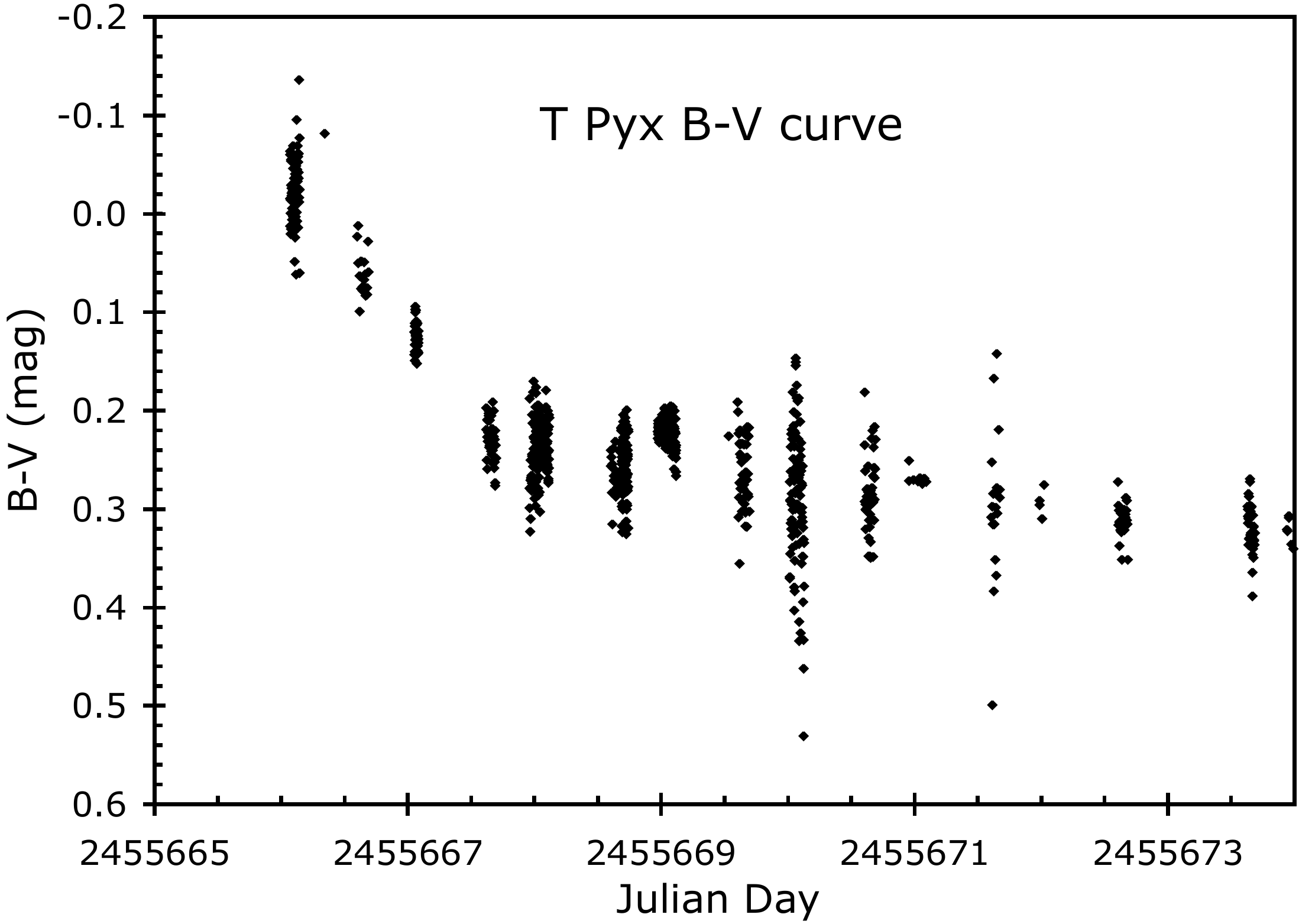}
\caption{
B-V color over initial rise.  We have the first and only measure of the colors of the initial rise of a nova.  The color does change somewhat over the first two days of the eruption (covered by the model fit in Figure 3), suggesting that the photospheric temperature does change over this interval, with this being contrary to the simple model of a uniform expanding shell with a constant temperature.  The initial colors ($B-V=-0.02$ and $V-I=+0.39$) are not those of a blackbody.  In addition, the SMARTS B-band magnitudes from JD 2455665.4752 imply $B-V = -1.0$ if the earliest moments of the light curve are smooth and monotonic in the rise.}
\end{figure}

\clearpage
\begin{figure}
\epsscale{0.8}
\plotone{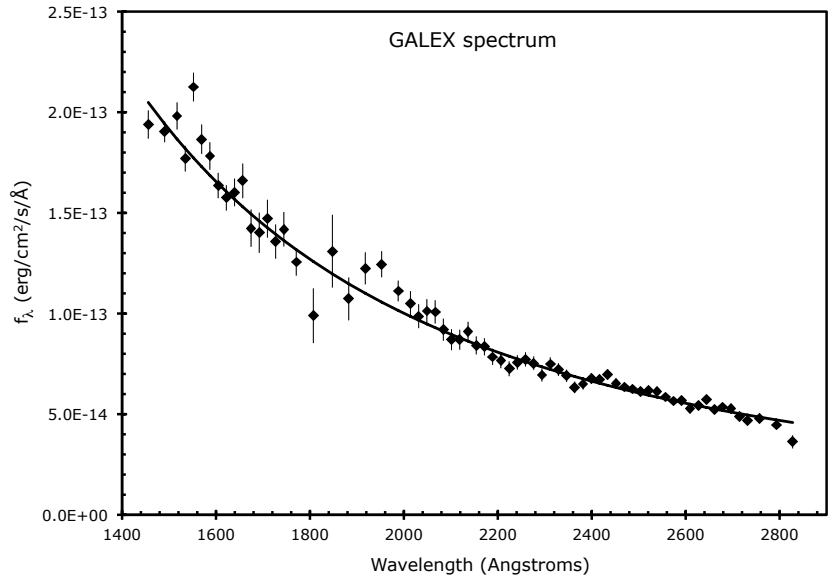}
\caption{
{\it GALEX} spectrum.  This spectrum of T Pyx in quiescence was taken in December 2005 with the {\it GALEX} satellite.  The error bars are moderately high around 1800-1900\AA due to that being the region where the NUV and FUV cameras join together.  The original spectrum showed a prominent broad dip centered on 2175\AA, due to the usual absorption by the interstellar medium, with our dereddening corrections designed to eliminate this dip.  The resultant spectrum is a good power law with an index of -2.25$\pm$0.03.  The C IV line at 1550\AA~is significant, and the He II line at 1640\AA~is likely visible but not significant.  This spectrum can be compared with the ultraviolet spectrum derived from greatly longer exposures with the {\it IUE} satellite that covers a somewhat broader wavelength range (Selvelli et al. 2010).}
\end{figure}

\clearpage
\begin{figure}
\epsscale{0.8}
\plotone{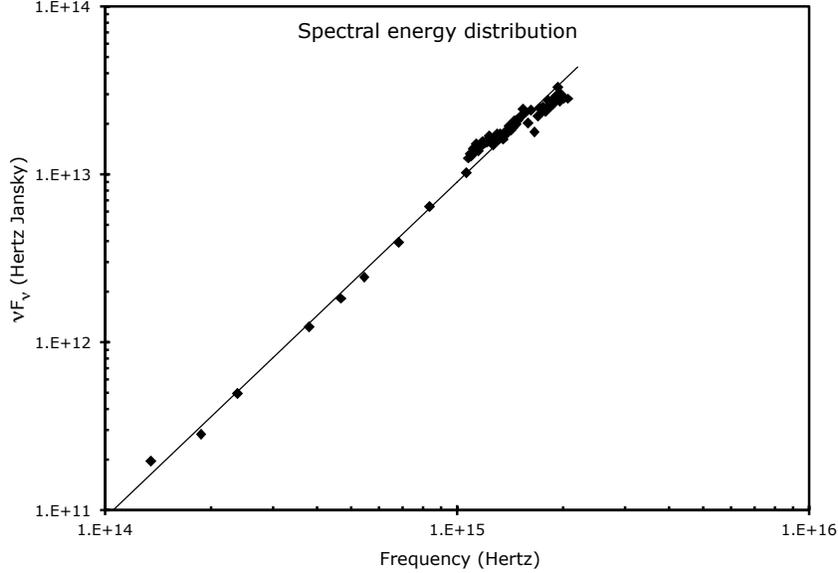}
\caption{
T Pyx spectral energy distribution.  This SED plot is constructed from the {\it GALEX} spectrum plus UBVRIJHK photometry in Schaefer (2010).  To a close approximation, the spectral energy distribution is a simple power law with $\nu f_{\nu} \propto \nu^2$ or $f_{\nu} \propto \nu$.  Within this broad overall shape, the ultraviolet portion of the spectrum is somewhat flatter, with a slope that goes as $f_{\nu} \propto \nu^{0.25}$.  This is completely different from a blackbody spectrum, as the Rayleigh-Jeans tail has $f_{\nu} \propto \nu^2$.  This is also completely different from a disk spectrum, which should go like $f_{\nu} \propto \nu^{1/3}$ only to for frequencies above the observed range (due to any disk requiring a temperature of the outer edge of 20,000$\degr$K or hotter), while any disk SED must have turned over to a Rayleigh-Jeans tail in most of the plotted spectral range.  There is one additional point that has not been plotted, and that is that the X-ray brightness of the central source can be represented by a point greatly below the plot at a frequency of $\sim 2 \times 10^{17}$ Hz.  This point forces the power law rise to high frequency to undergo some sort of a cutoff between $2 \times 10^{15}$ and $2 \times 10^{17}$ Hz.  Unfortunately, there is no way to better constrain the cutoff, and this leads to a factor of 10,000$\times$ uncertainty in the accretion energy.  }
\end{figure}

\clearpage
\begin{figure}
\epsscale{0.6}
\plotone{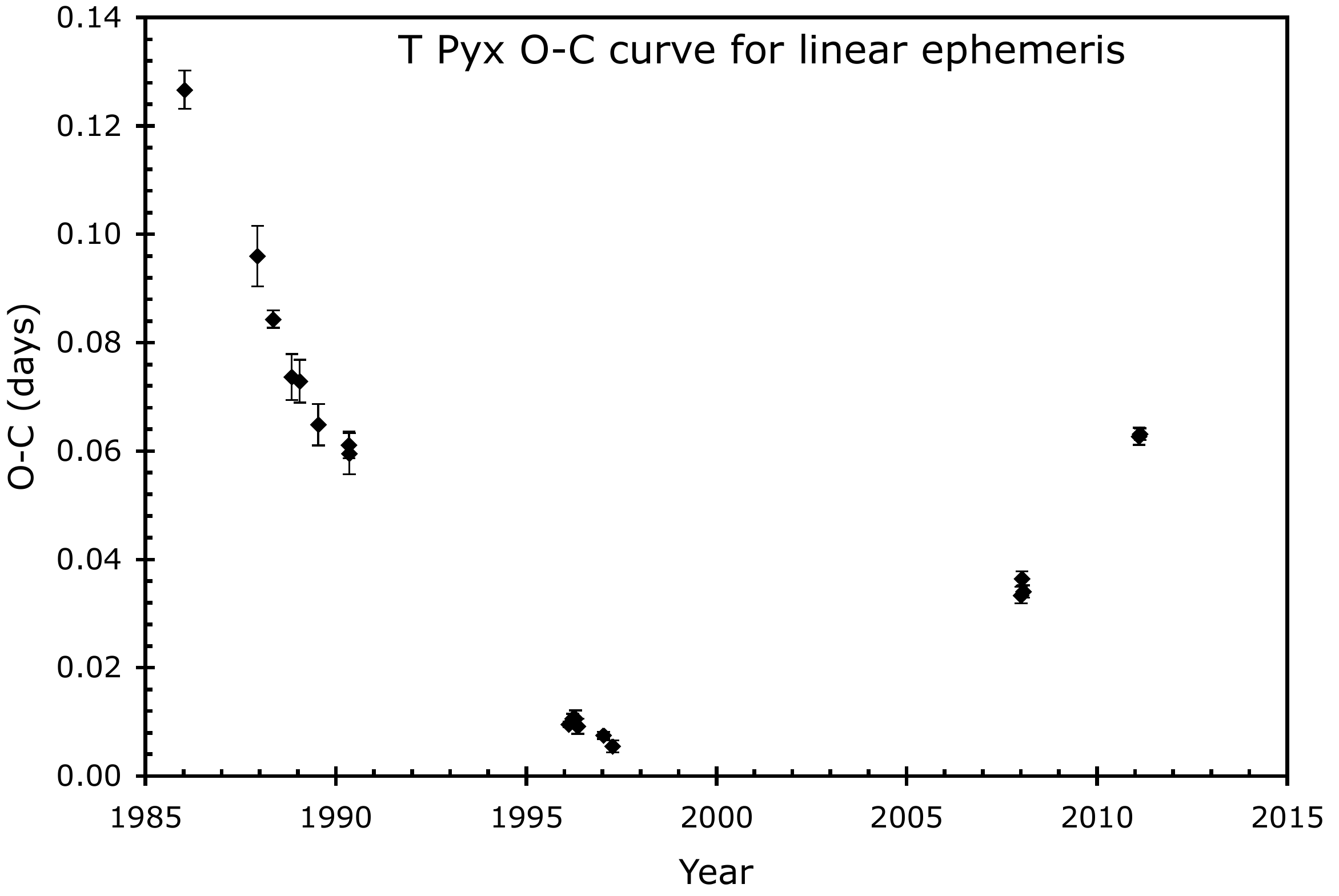}
\plotone{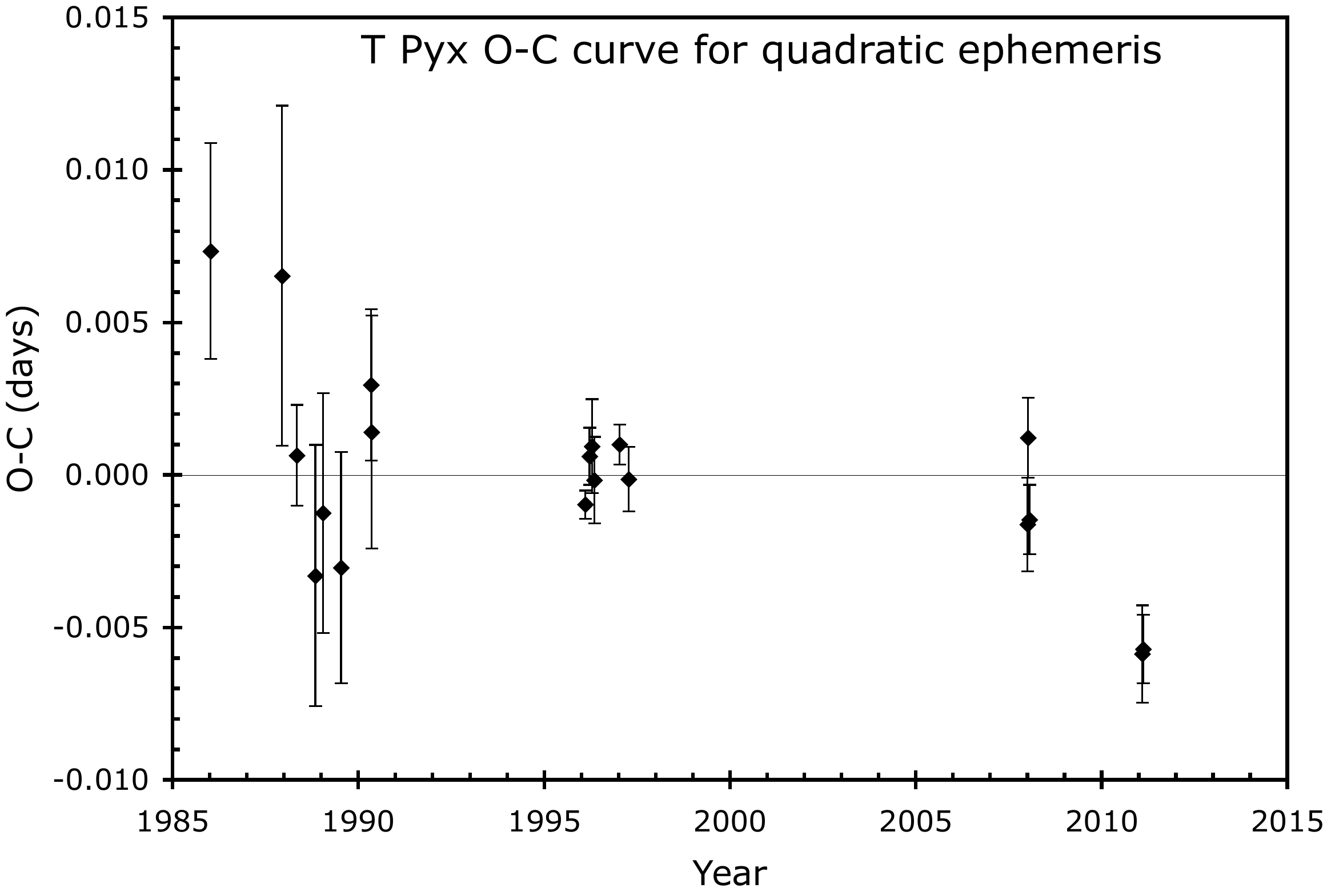}
\caption{
T Pyx O-C curve.  T Pyx has a stable periodic modulation in its light curve, and we have minimum times from 1986 until February 2011 (just 48 days before the eruption).  The upper panel shows the O-C curve for a linear ephemeris (Equation 2), in which we see a closely parabolic shape.  This indicates that T Pyx has a steady change in orbital period, just as expected for steady conservative mass transfer at a rate of $1-7-3.5\times 10^{-7}$ M$_{\odot}$ yr$^{-1}$.  The lower panel shows the O-C curve for the predicted times coming from the quadratic ephemeris (see Equation 1) of Uthas et al. (2010).  This steady period change well describes the minimum times from 1988 to 2008.  But the two minimum times from February 2011 are consistent and highly significantly below the predictions from a single parabola, so apparently T Pyx underwent some sort of a small $\dot{P}$ change sometime around 2008, such that its orbital period just before the 2011 eruption is slightly different (7.5 parts per million) from the period given by Equation 1.  Such a change in $\dot{P}$ is expected due to the observed 18\% decrease in system brightness (and hence a 18\% decrease in accretion rate) from 2004.5-2011.}
\end{figure}

\end{document}